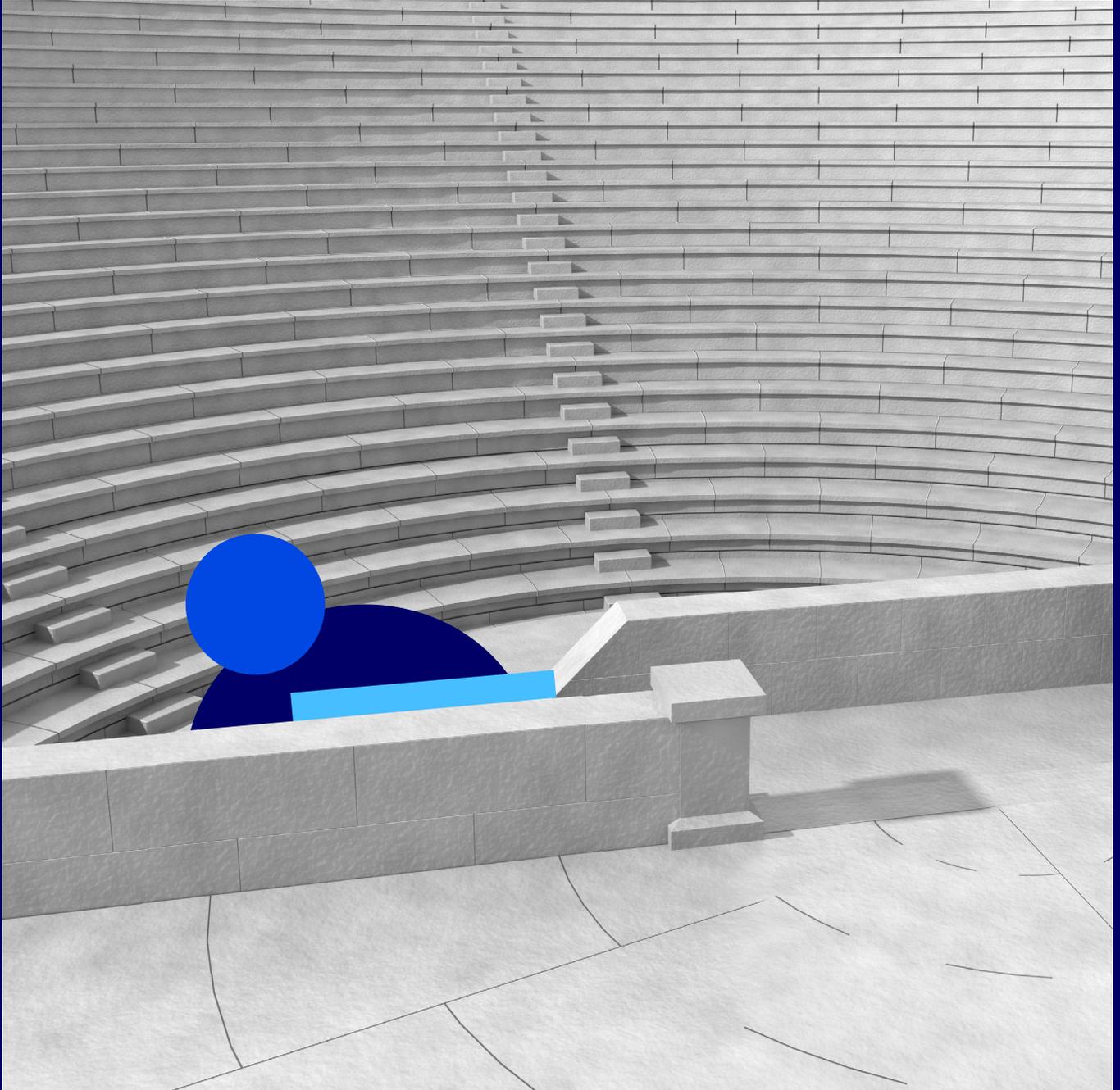

# APPLICATIONS OF
# ARTIFICIAL INTELLIGENCE
# TOOLS TO ENHANCE
# LEGISLATIVE ENGAGEMENT:
## CASE STUDIES FROM MAKE.ORG AND MAPLE

**SEPTEMBER
2024**

# CONTENTS

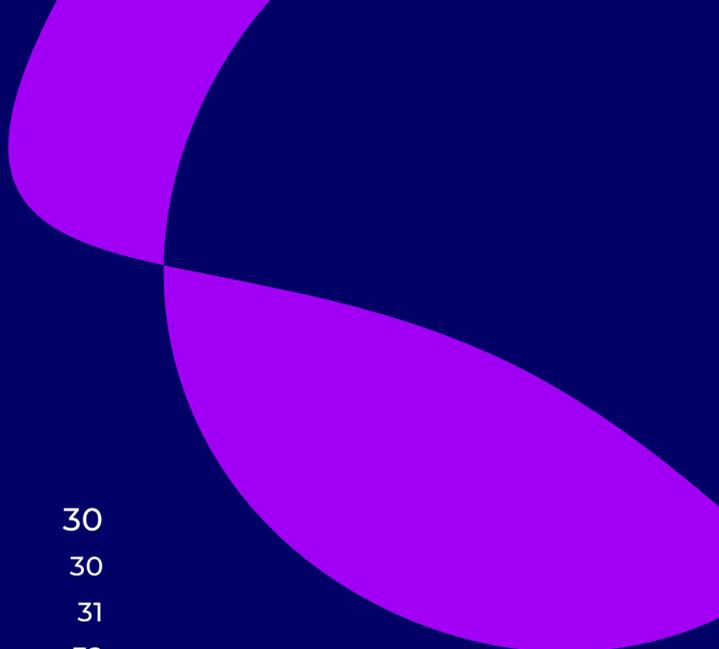

# ABSTRACT

This paper is a collaboration between Make.org and the Massachusetts Platform for Legislative Engagement (MAPLE), two non-partisan civic technology organizations building novel AI deployments to improve democratic capacity. Make.org, a civic innovator in Europe, is developing massive online participative platforms that can engage hundreds of thousands or even millions of participants. Their latest platform, Panoramic, is an AI-powered platform used for the first time to enable citizens to engage with the debates of the 2023 Citizen Assembly on End of Life in France. MAPLE, a volunteer-led NGO in the United States, is creating an open-source platform to help constituents understand and engage more effectively with the state law-making process.

**We believe that assistive integrations of AI can meaningfully impact the equity, efficiency, and accessibility of democratic legislating.** We draw generalizable lessons from our experience in designing, building, and operating civic engagement platforms with AI integrations. We discuss four dimensions of legislative engagement that benefit from AI integrations: (1) making information accessible, (2) facilitating expression, (3) supporting deliberation, and (4) synthesizing insights. We present learnings from current, in development, and contemplated AI-powered features, such as summarizing and organizing policy information, supporting users in articulating their perspectives, and synthesizing consensus and controversy in public opinion.

We outline what challenges needed to be overcome to deploy these tools equitably and discuss how Make.org and MAPLE have implemented and iteratively improved those concepts to make citizen assemblies and policymaking more participatory and responsive. We compare and contrast the approaches of Make.org and MAPLE, as well as how jurisdictional differences alter the risks and opportunities for AI deployments seeking to improve democracy. We conclude with recommendations for governments and NGOs interested in enhancing legislative engagement. Our principal recommendation is to encourage thoughtful experimentation using AI to enhance participatory technology with an emphasis on approaches that build trust with potential user communities.

**WRITTEN BY**
ALICIA COMBAZ, MAKE.ORG
DAVID MAS, MAKE.ORG
NATHAN SANDERS, MAPLE
MATTHEW VICTOR, MAPLE

# INTRODUCTION

Democracies around the world face dueling crises of eroding trust in civic processes and rising polarization and vitriol, both fueled in part by modern technologies such as social media.

> **Technology is not neutral, because it is designed and applied within the moral and cultural context of its developers. For those focused on enhancing public participation in governance and the resilience of democracy, there is an urgent need to affirmatively develop technologies supportive of democratic processes as a counterweight to the societal factors, including other technologies, which afflict them.**

Generative AI systems are capable of instantaneously summarizing and explaining diverse corpuses of text. This has enormous applications in education, business, and science, but it also can benefit democracy. Applied properly, these capabilities can be a force for increasing engagement in the legislative process and making constituents more informed about and active in policymaking, while simultaneously making legislators more responsive and connected to their constituents.

Public engagement and deliberation is a core function of legislative branches of democracies (Curato et al., 2017). Yet there is relatively little development of conceptual frameworks for enhancing engagement, measuring engagement, and understanding its impact (Leston-Bandeira & Siefken, 2023), particularly in contrast to electoral politics. Much of the attention on public engagement with legislation today focuses on social media platforms. Researchers have characterized how social media is a primary news source for many (Wang & Forman-Katz, 2024), a significant channel for civic engagement (Vandermaas-Peeler et al., 2018), and subjects users to intense polarization and echo chambers (Barrett, Hendrix, & Sims, 2021). Since social media has failed to create a space for peaceful and constructive debate, civic energy would be more effectively channeled through digital infrastructure purposefully designed to serve public interests.

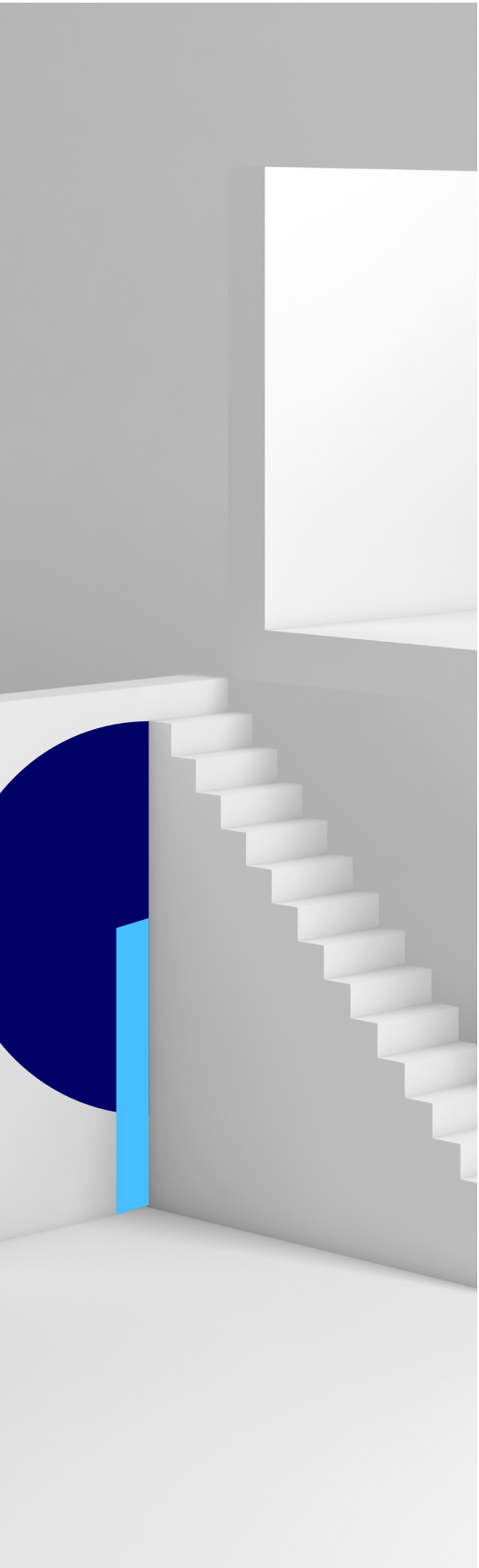

There is a small but growing field of literature exploring how AI tools can facilitate, mediate, and broaden democratic engagement. Much of this work focuses on the executive rather than the legislative branch. Chen et al. (2023) analyzed present-day uses of AI chatbots across twenty-two US state agencies, such as providing alternative communication channels and helpdesk services for agency clients with 24/7 availability. Likewise, Cortés-Cediel et al. (2023) analyzed chatbots in use at the local, regional, and national level in Spain and developed a conceptual framework for evaluating their impact on citizen participation. Fewer works have focused on AI applications to legislative engagement. Kreps & Jakesch (2023) used a large language model (GPT-3) to test constituents' response to communications from legislators that were written by, or with assistance from, AI. They reported that constituents responded most favorably to AI generated messages that had oversight by humans, and preferred these messages to both boilerplate messages and individual, human-written replies. Researchers, including Fish et al. (2023) and Dai et al. (2024), have developed formal methods for using AI models to scalably guide large numbers of human stakeholders towards optimal consensus positions, suggesting future methods for AI-assisted legislative deliberation that more broadly engage the public.

Putting this research to practice, there is a growing number of initiatives deploying AI-driven tools to improve democratic engagement. Over decades, civic technologists have accumulated a wealth of knowledge regarding proper design, product and go-to-market decisions[1], which provides a foundation for the nascent AI-driven civic tools being developed. Gesnouin et al. (2024) developed LLaMandment, a large language model for summarizing French legislative proposals aimed, in part, at making the legislative process more accessible to citizens, journalists, and the broader public. Other initiatives, such as "Talk to the City", are demonstrating the value of LLM-based collective decision-making tools in a variety of settings, including labor unions, NGOs and decentralized autonomous organizations (DAOs).[2]

We are investing in these implementations and demonstrations of AI assistive features because we believe they can meaningfully impact the equity, efficiency, and accessibility of policymaking within democracies. In many jurisdictions, legislature websites are the only source for policy information, and such sites are often complicated to navigate, display inaccessible legislative verbiage, and require familiarity with legislative processes. However, we understand the limitations of a techno-solutionist paradigm and embrace a multidisciplinary approach to integrating technology with society (Lindgren & Dignum 2023). We have therefore designed and tested the tools described in this article with intensive involvement of stakeholders across society, including citizens, scholars from diverse fields, advocacy organizations, and policymakers. We hope that our examples can help others to steer towards human-centered integrations of this technology and we will discuss the risks and limitations of the applications we are developing.

---

1    https://civictech.guide/

2    https://www.ie.edu/cgc/news-and-events/news/using-ai-to-inform-policymaking-what-we-can-learn-from-3-use-cases-of-talk-to-the-city/

> To be effective tools for advancing the democratic principles of equitable access, transparency, and civil discourse, AI systems must be deployed responsibly, with disclosure about their scope of use and limitations, in ways accessible to as many users as possible, and with assurances of fair treatment.

We conclude with recommendations for policymakers and civic organizations based on the experiences of our organizations with innovative deployments of AI to legislative engagement.

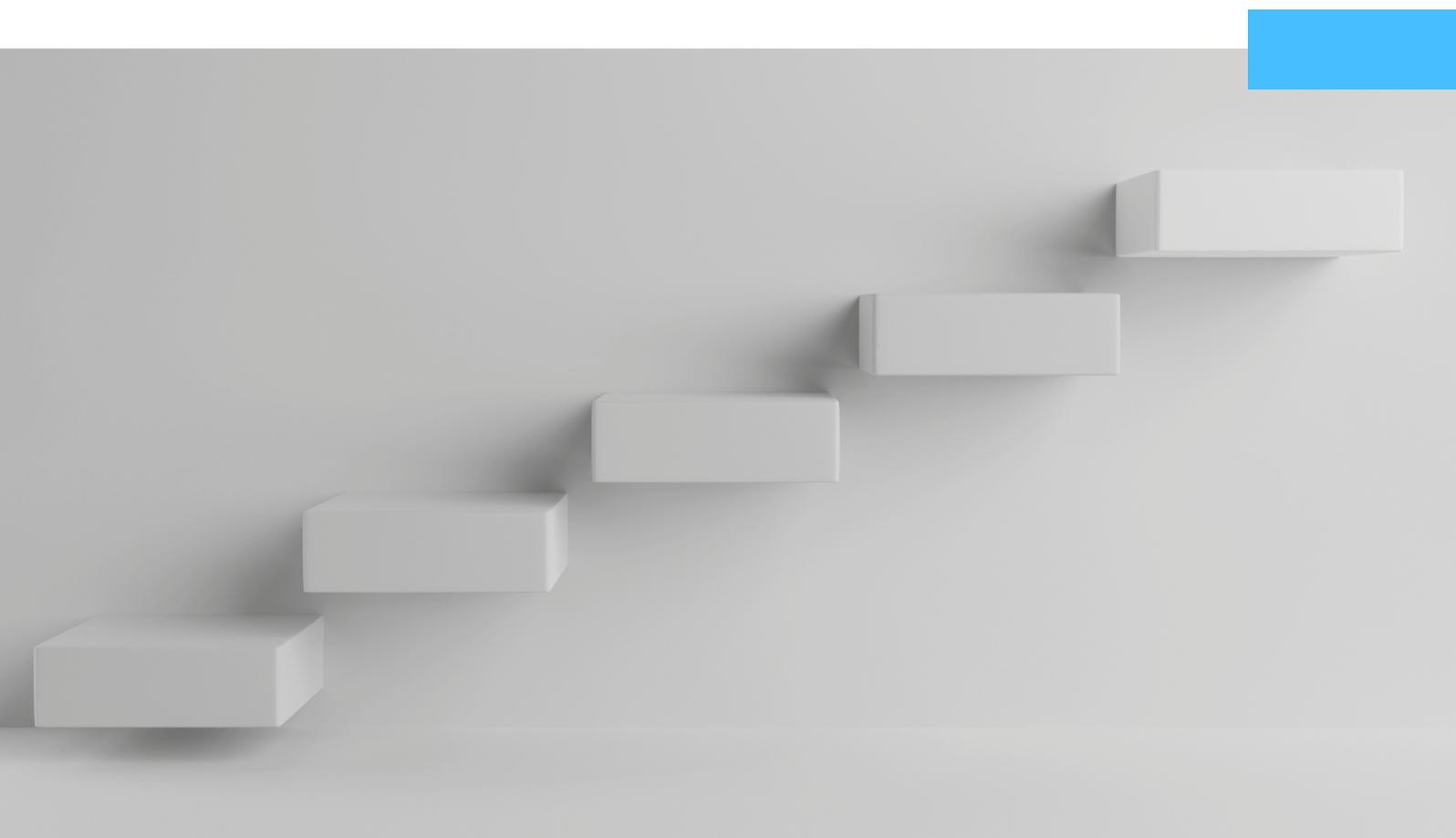

# ORGANIZATIONAL CONTEXT OF MAKE.ORG AND MAPLE

This section introduces the two civic organizations discussed in this paper and their civic context, and outlines the goals behind the AI integrations they are developing.

## ◼ ABOUT MAKE.ORG

**Make.org is a nonpartisan, independent organization committed to empowering citizens and mobilizing all of civil society to bring about positive societal change.**

Make.org believes that such changes can only be achieved through popular consensus and has developed a suite of pioneering tools to engage millions of people in the project of collective decision-making. Their approach is to identify the most widely supported ideas across a society, and then build collective action that is legitimized by the adhesion of the greatest number of people. Over the past 8 years, Make.org has engaged more than 10 million citizens, more than 50 institutions and more than 1,000 associations and partners in Europe.

Make.org's Consultation Platform allows for engagement around an open question; citizens can make proposals and vote on the proposals of others. The Consultation Platform is instrumental in creating the Agenda of Hope[3] in the run-up to the European elections, with over 1.5 million votes cast and more than 5,000 proposals submitted. This agenda is complemented by the response of all European political parties to each of the citizens' priorities. Unprecedented dialogues between citizens, who are widely involved, and their representatives are being created.

Another participatory platform, Dialog, brings stakeholders together to collaborate and design impactful projects. The French Ministry of Economy has used this tool to engage citizens and gather their feedback on a variety of policy proposals. This platform is also used by the Bertelsmann Foundation and the German Ministry of Interior as part of the "Forum Against Fakes" project[4] to connect participants of a citizens' convention and German citizens. Finally, Panoramic, a native AI platform makes it easier for everyone to access complex content such as the expansive deliberations within citizen assemblies.

---

3    https://eurhope.org/en

Make.org launched the Democratic Shield[5] initiative at the end of 2023. The initiative united a set of actions for institutions and civil society, aimed at safeguarding the integrity of the European elections. Over the past few months, an initiative Task Force has mobilized and engaged with the public and institutions to implement these measures, fortifying European democracy ahead of the European elections.

Finally, in April 2024, Make.org announced the launch of a 2-year research program, in partnership with Sciences Po, Sorbonne University and the CNRS: the 'Democratic Commons[6]'—the first AI global research program to reinforce democracy. This project has attracted the world's leading experts in ethical AI: Hugging Face, Mozilla.ai, Aspen Institute, Project Liberty Institute, and Genci. It will bring together over fifty researchers and engineers. Its primary objective is to develop and share a social science scientific framework for determining democratic principles applied to AI, a model for evaluating the biases of large language models (LLMs) against these principles, debiased LLMs, and citizen participation platforms that adhere to these principles.

## ◼ ABOUT MAPLE

**The Massachusetts Platform for Legislative Engagement (MAPLE) is an independent, nonpartisan, nonprofit organization building web tools to enhance public engagement with the Massachusetts (MA) legislature.**

The primary features of the MAPLE web platform are (1) educational materials about the legislative process and public participation, (2) a searchable database of current MA bills and their status and history, (3) a repository of public comment on each bill, consisting of testimonies submitted through the platform, (4) a function for users, individuals and organizations, to submit their own testimony on any bill, and (5) profile pages (optional for individuals) with an aggregated view of testimonies from each organization or individual user. AI integrations are under development to extend each of these features. MAPLE's open-source codebase makes it possible to clone and adapt these features to any other jurisdiction or community.

---

4   https://about.make.org/articles-en/forum-against-fakes-a-broad-online-citizen-participation-to-support-the-work-of-a-new-citizens-panel-in-germany-on-how-to-tackle-misinformation

5   https://about.make.org/articles-en/election-protection-on-the-eu-agenda-the-democratic-shield-initiative

6   https://about.make.org/articles-en/presentation-of-the-democratic-commons-global-research-program-at-vivatech-for-ethical-ai-in-service-of-democratic-resilience

MAPLE's strategy is to improve the democratic process by providing a safe, widely accessible space for sustained and productive public discourse on public policy. Unlike social media, MAPLE provides a convenient yet impactful channel for civic engagement by providing a moderated space for any constituent to seek out and contribute relevant information on actionable matters. All MAPLE submissions are reviewed by human moderators (possible due to the platform's narrow scope and limited scale). Users are provided self-curation tools to sort and filter content, which is displayed without any further algorithmic adjustments. There are no comment sections on posted testimony nor follower counts shown. Together, these design decisions reduce incentives and opportunities for reductive and performative behavior and foster a shared reality centered on community stakeholder perspectives and proposed legislative changes.

Massachusetts is a particularly interesting context in which to study the use of digital tools and AI for facilitating legislative engagement. Relative to other US states, MA state-level politics has exceptionally low levels of democratic participation and transparency. MA legislature elections are among the least competitive in the country (Lannan, 2022). In the 2022 election, almost 60% of the 200 winning candidates faced neither a primary nor general challenger. Meanwhile, MA is one of only four US states that have exempted their legislatures from public records laws (Gomez, 2018)—meaning that there is no guaranteed public access to legislative documents, such as testimony. This lack of transparency represents a barrier to public understanding of the legislature's reasoning, and to whom the legislature is listening and responding, when making decisions on public policy. Nonetheless, researchers have in some cases produced datasets of testimony from MA legislative committees, demonstrating how illuminating these materials can be to understanding channels of democratic influence (Culhane et al., 2021).

**The value proposition for MAPLE users is an easier pathway to having their voice heard in public policymaking.**

The MAPLE platform helps users who would not otherwise be able to submit written testimony to the legislature to do so, and also provides them with a public platform to share and attract attention to their perspective. Even users who do not choose to publish testimony on MAPLE can benefit from its use as a research tool, to understand legislation and public opinion about policy proposals, and as an assistive platform to help them develop and direct testimony to the legislature.

MAPLE produces societal benefits, as well. MAPLE seeks to increase MA democratic participation, specifically legislative engagement, by making the process for constituents to research and submit public comment on legislation through official legislative channels easier. MAPLE also seeks to strengthen constituent connections to local civically active organizations, which are traditional sources of political information, coordination, and mobilization (Li & Zhang, 2017). Finally, MAPLE seeks to increase the transparency and accountability of the MA legislature by creating the first public repository of written testimony in Massachusetts.

The organization is led by volunteers affiliated with Boston College, Northeastern University, and Harvard University, and was incubated by Northeastern Law School's NuLawLab and Code for Boston. MAPLE has been in development since 2021 and was publicly launched in April 2023. In its first year, the platform conveyed and archived 480 testimonies from seventy-eight individual users and civic organizations. The platform is supported financially by individual and foundation contributions, and is an initiative of the US nonprofit organization Partners in Democracy Education[7].

---

7    https://partnersindemocracy.us

# ENHANCING DEMOCRACY ACROSS FOUR DIMENSIONS

**In this paper, we describe two novel projects that demonstrate the potential for generative AI to help constituents understand legislative proposals and policy issues, articulate their own preferences, bridge language barriers, and find consensus among diverse ideas.**

We discuss the AI integrations of these projects, including features already deployed and others in development, with respect to four dimensions for improving the quality, process and outcomes of our civic engagement and law-making structures. We describe the innovations of these projects in each area: (1) making it easier to understand complex materials; (2) helping individuals express themselves; (3) facilitating effective deliberation and identifying consensus; and (4) conveying insights and consensus to decision-makers.

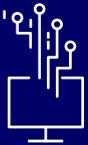

## DIMENSION #1

## MAKING INFORMATION ACCESSIBLE

The digital age has brought an unparalleled wealth of information to our fingertips, but without sufficient mechanisms to alleviate the information overload that harms decision-making and inhibits constituent participation (Hołyst et al., 2024). This problem is particularly acute in the law-making context, as the complex and often-antiquated structures of legislatures present their own accessibility issues for the general public. Our democratic institutions will benefit from structures that enhance our understanding of one another and of legislative proposals.

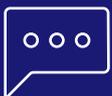

## DIMENSION #2

## FACILITATING EXPRESSION

Furnished with quality, relevant information, individuals are better prepared to contribute to public discussion. However, such participation is in itself a daunting task for many who may fear embarrassment or ostracism for expressing a poorly informed or socially disfavored opinion (Weeks, Halversen, and Neubaum, 2024). Others may feel uncomfortable with the perceived quality of their public participation, particularly if they lack formal education or knowledge of local languages and customs.

Digital civic infrastructure can be designed to mitigate these barriers to participation. For example, Make.org structures focused conversations in smaller settings to allow for more comfortable expression and sincere discussions, while MAPLE will provide tools to help constituents draft more relevant, evidence-based testimonies and to overcome the intimidation often felt when submitting formal public testimony.

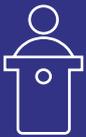

**DIMENSION #3**

## SUPPORTING DELIBERATION

Deliberation is, in itself, a public good, as it helps people understand and engage with alternative viewpoints. Ideally, deliberations can illuminate consensus that turns into actionable directives for policy-makers. How those deliberations occur and are reviewed and evaluated are of immense significance. On social media, interactions are structured to increase engagement, and therefore algorithms often reward antagonistic and outrageous behavior (Munn 2020) and high numbers of "reshares" (often of the most reductive yet engaging material) present the only indication of consensus. But civic-oriented platforms can be built with other objectives in mind, such as eliciting sincere expressions, orienting users to be open to new ideas, and building trust. AI can be leveraged to achieve these goals, facilitating better moderated public deliberation and consensus building.

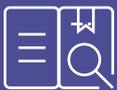

**DIMENSION #4**

## SYNTHESIZING INSIGHTS

Emergent consensus is a "tree fallen in the woods" unless it is effectively conveyed to policymakers. The effects of "information overload" do not spare legislators, and combined with the often under-resourced condition of many (particularly local) policymakers, there are many challenges to conveying a consensus in a manner that will beget serious attention from legislative bodies. How does a legislator know whether a consensus is soundly reached, or represents input from all stakeholders? How does a legislator actually take action on a consensus, even if the public directive is clear?

To address these issues, Make.org presents several tools each geared towards the different conditions and sizes of deliberations—including projects initiated by government actors. MAPLE's features fit into existing processes of policymaking to convey actionable consensus to legislators. For both organizations, the degree to which policymakers are engaged in or open to considering the outputs of the deliberative process is a critical factor.

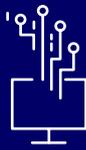

# DIMENSION #1

## MAKING INFORMATION ACCESSIBLE

Assemblies, and particularly citizens' conventions, are high-quality democratic events. In France, the Economic, Social, and Environmental Council (CESE), the third constitutional assembly of the Republic, is an essential component in French democracy and a key institutional actor in citizen participation.

The CESE publishes recommendations to the government and parliament and participates in the development and evaluation of public policies in its fields of competence. The CESE brings together 175 members, men and women from the field, appointed by intermediate bodies: associations, unions of employees, employers' organizations, and so on. Since 2021, the Council has been officially entrusted with new missions, allowing citizen participation to enrich its work in a useful way. It is in this context that the CESE organized a citizens' convention on the end of life, including the topics of euthanasia and assisted suicide, between September 2023 and April 2024.

### CITIZEN'S CONVENTION ON END-OF-LIFE ISSUES

Announced by the French President of the Republic on September 13, 2023, this Citizens' Convention brought together citizens whose work was intended to shed light on the following question: "Is the framework for end-of-life care adapted to the different situations encountered or should any changes be introduced?" During nine working sessions, the participants were tasked with deepening the aspects of this question to build dialogue, debate, and finally sketch out perspectives and consensus. The Convention was nourished by the expertise and experience of all stakeholders, including professionals such as palliative care teams who are regularly confronted with the end-of-life situations in their practice and daily lives. At the end of nine working sessions and twenty-seven days of debate, the Citizens' Convention presented its conclusions and adopted its final report.[8]

---

8   https://www.lecese.fr/sites/default/files/documents/CCFV/Conventioncitoyenne_findevie_Synth%C3%A8se.pdf

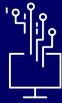

These conclusions and this report made it possible to prepare the examination of the bill on the end of life that began in the National Assembly in May 2024.

The CESE is particularly transparent in all of its work; reports and videos are posted online on its website or social networks. However, it is clear that this "unaccompanied" transparency seems ultimately insufficient. In citizens' convention on the end of life, it appears that citizens face two complexities: the length of the content (no less than forty hours of discussion, reports averaging more than ninety pages), and also the technical nature of the subjects discussed; few people have the time, energy and ability to delve into this work, which is nevertheless key to our democracies;

## ABOUT PANORAMIC

Within this context, Make.org, partnered with the CESE to develop a solution that uses the power of generative AI to make the content of this convention available in a simple and accessible way. The principle is simple: collect the content that can be shared, moderate it, and thanks to Retrieval-Augmented Generation (RAG; Lewis et al., 2020) technology, any citizen can ask a question to the interface and get a simple answer. The RAG technology combines two components. The first component is a text embedding model powering a semantic search engine. It analyzes the user's query to understand the intent and meaning behind the search. The second retrieval system can then search through a vast amount of documentation. The retrieved documents are then fed to the LLM. This additional information helps the LLM to refine its initial response.

> **The aim is to involve people who are far removed from the subject of politics, those who express themselves little or who feel little concerned. It is by this yardstick that the proposed experience is truly democratic and inclusive.**

That's why the proposed interfaces of Panoramic are very simple; Make.org invites citizens to take very simple initial actions, which will lead to progressively more complex responses. Three entry points have been determined: the theme, the suggested prompt, and the open prompt (Figure 1).

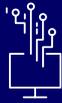

CASE STUDY #1
**MAKE.ORG**

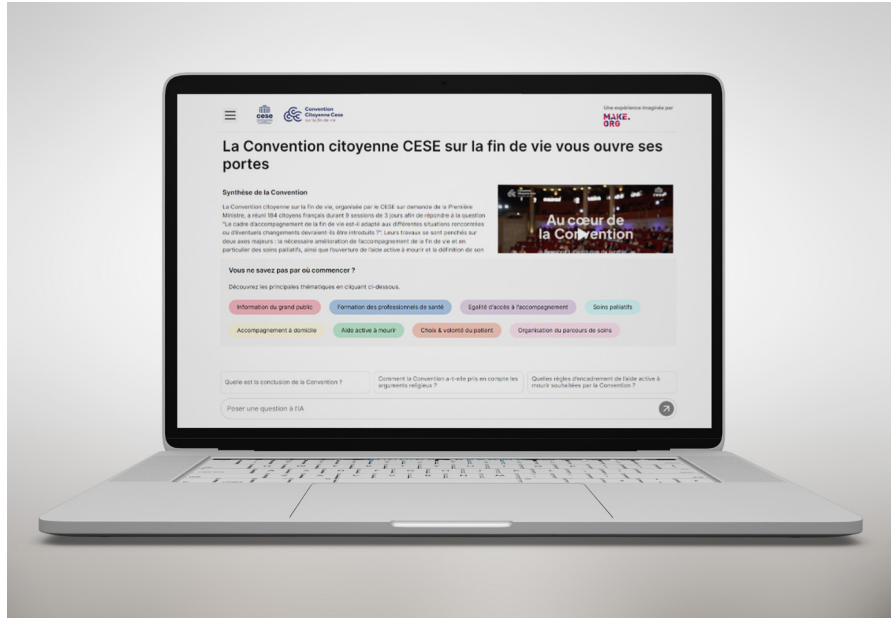

**FIGURE 1.**
This image depicts the three entry points on the *Make.org* Panoramic Home page.

Each answer is composed of an easy-to-understand text and the sources that built the answer; you can also go and watch the exact moment of the subject that matters to the user (Figure 2).

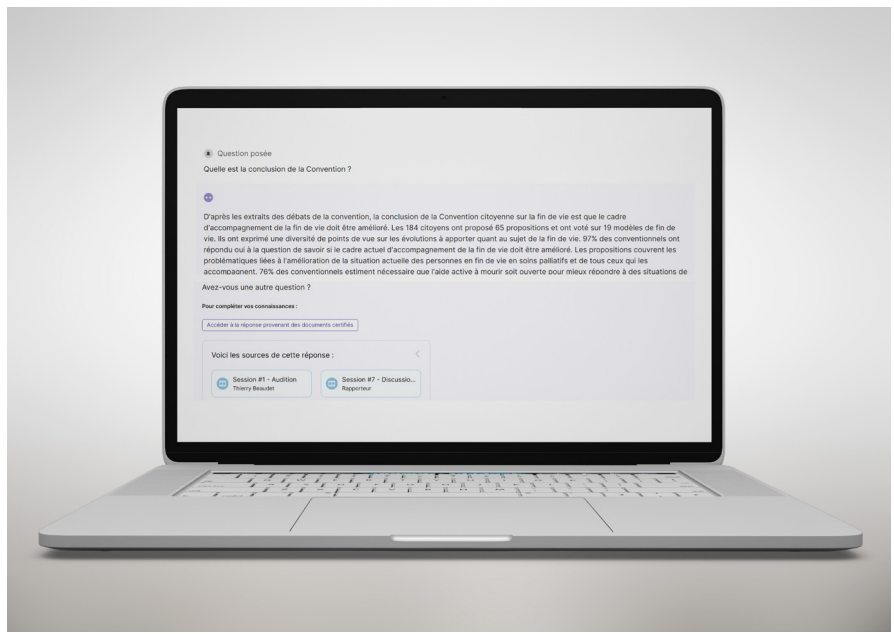

**FIGURE 2.**
Make.org Panoramic screenshot depicting an extract from the written answer, and the link to the sources.

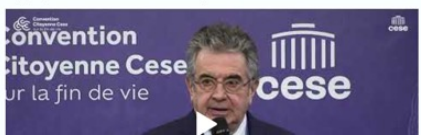

**FIGURE 3.**
Depicting the video source description on the Make.org Panoramic platform, with the exact timestamp.

The interface is set to evolve over the coming months. To maximize citizen engagement with the platform, we focus on improving specific key performance indicators (KPIs). We prioritize KPIs that measure active interactions from citizens, as opposed to more passive metrics like time spent on the platform. This choice reflects our commitment to empowering citizens to become active participants in the democratic process.

One KPI is the first-visit contribution rate; that is, the ratio of the number of people who make a first click versus the number of people who arrive on the platform. This is currently 40% and should approach 65% after iterative improvements to website copy (clarity of the promise), UI (hierarchization of information), and UX (work on the "articulation barrier," for example). The second KPI is the second-click rate; that is, the percentage of users who engage in a second action among those who have taken one. This metric will improve over time after optimizing the length of the AI response, its design, its content (prompt engineering), and suggested user actions.

Building an interface or crafting a unique experience is just the first step. To truly connect and engage potential visitors, Make.org seeks to combine these elements with strategic social media campaigns. The messages are simple, clear and focused on a theme and issue that is very tangible for citizens; it's by associating this message with a simple, engaging experience that Make.org reaches people who are rarely involved with civic deliberation (Figure 4) .

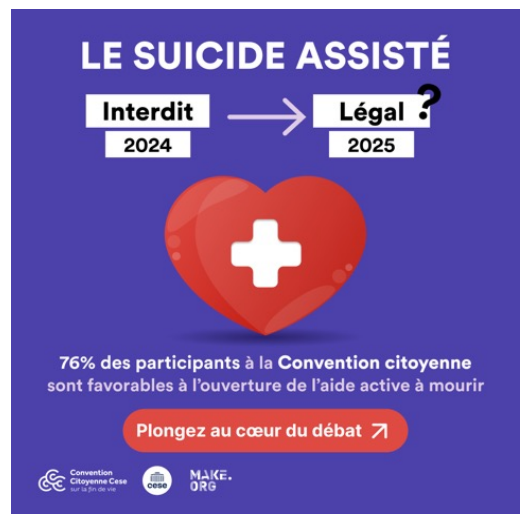

**FIGURE 4.**
Depicting an acquisition campaign visual for the Make.org Panoramic platform.

CASE STUDY #1
**MAKE.ORG**

## LESSONS LEARNED FROM PANORAMIC

While experimenting to improve the quality of the AI systems, the Make.org team learned some lessons on maximizing outputs of generative AI and avoiding the biggest pitfalls.

■ **Data quality:** The quality of data fed to the AI was crucial for achieving accurate results. This means not only ensuring high-quality transcripts, but also providing ample context to help the AI's understanding of the information. For instance, Make.org invested significant time in identifying each speaker and enriching the data with details about the debate's context.

■ **Prompt engineering:** As with any AI project, crafting effective prompts is paramount for success. This balancing act is familiar to all AI practitioners: providing too much detail can overwhelm the model, while insufficient guidance can lead to inaccurate or fabricated outputs, a phenomenon known as hallucination. To ensure our prompts struck the right balance, Make.org implemented a rigorous evaluation process to identify the most effective formulations and mitigate the risk of hallucination. In particular, one effective method to avoid hallucination is to instruct the AI via a prompt to rely strictly on information derived from semantic search. However, there are instances where it can be beneficial for the AI to supplement the response with its general knowledge, such as when explaining technical concepts. We adjust the latitude given to the AI based on the topic at hand. For example, we provide strict guidelines when providing advice on cancer treatment but allow more flexibility for summarizing events like a tech conference such as Vivatech.

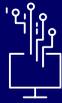

■ **Quality evaluation processes:** For optimal quality, a robust evaluation process is essential. While automated metrics like RAGAs[9] offer valuable insights, human evaluation remains unparalleled in ensuring the system aligns with citizens' needs. Therefore, Make.org implemented a hybrid approach, leveraging automated metrics to identify significant errors and regressions, followed by human evaluation for fine-tuning and optimization.

■ **Data source integration:** Our experience revealed that citizen inquiries extended beyond the debates themselves, encompassing broader topics and even the organization of the conventions. For example, citizens might ask "What is the CESE?" or "How were participants selected?" As these questions wouldn't be directly addressed within the citizen debates, Make.org proactively enriched the AI's data corpus with supplemental information to ensure it could comprehensively address such citizen inquiries.

Panoramic will be soon deployed at new assemblies—including parliamentary assemblies—throughout Europe. It can be used to follow and interact with any assembly, elected or not, such as a municipal council or a citizen assembly.

---

9    https://docs.ragas.io/en/stable/concepts/metrics/index.html

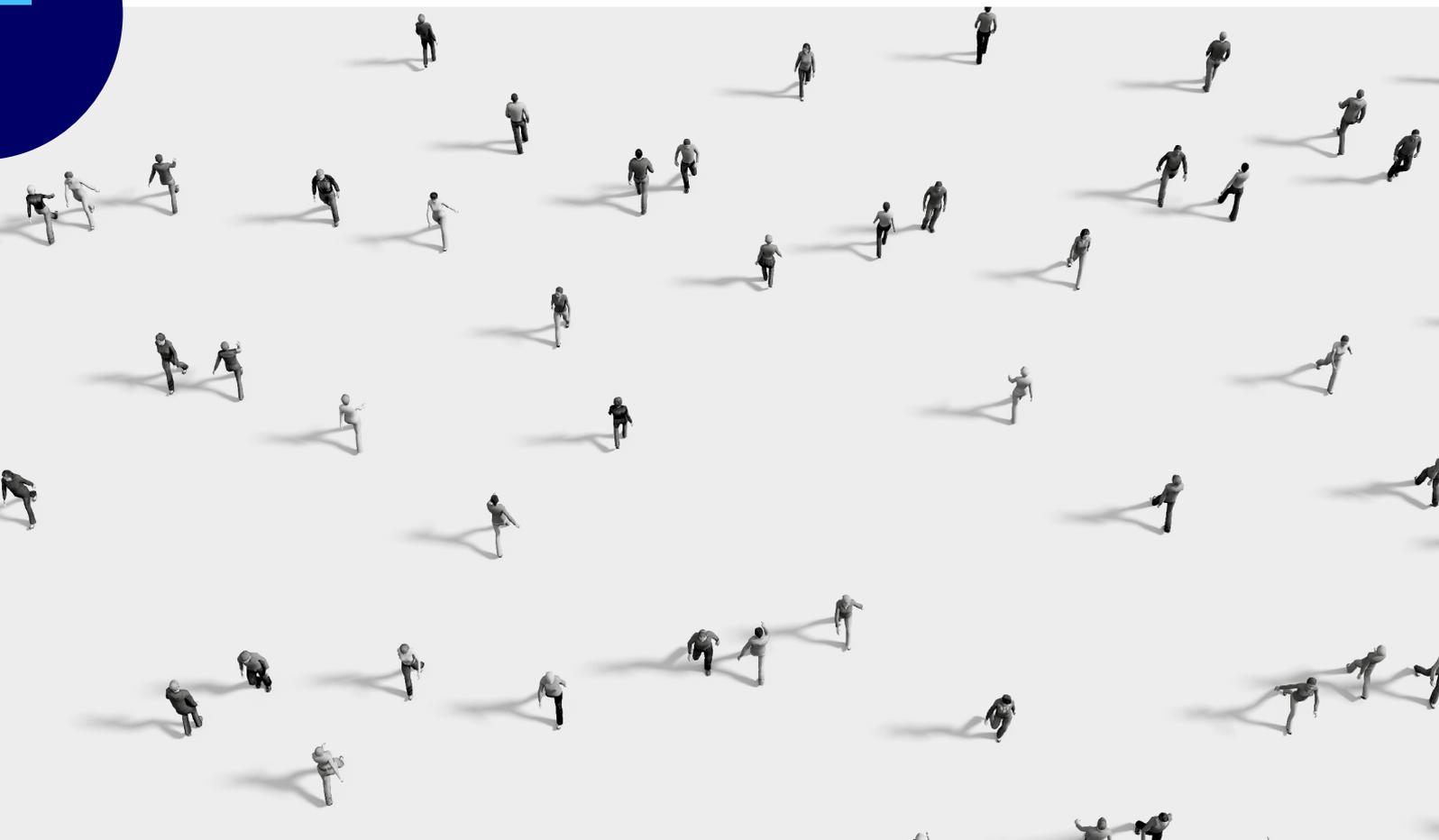

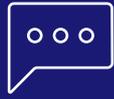

# DIMENSION #2

## FACILITATING EXPRESSION

Generative AI helps enable citizens of all knowledge and expression levels to participate in the decision-making process in four ways,

**1.** Breaking down language barriers: Generative AI can translate citizen input in real-time, allowing for participation regardless of native language.

**2.** Empowering brainstorming: AI can analyze large datasets and suggest new ideas or connections, sparking more creative discussions.

**3.** Overcoming hesitations: AI-powered tools can facilitate anonymous or voice-activated participation, encouraging those hesitant to speak up directly.

**4.** Improving articulation: AI can suggest phrasing or help articulate complex ideas, ensuring everyone has the tools to effectively express themselves.

Here are a few examples from the Panoramic platform that reflect some of these principles.

The vision behind Panoramic is an interface connecting a "mini public" with a "maxi public."

- A mini public is a small, diverse group of randomly selected citizens representing the broader population. They engage in deliberative discussions on specific issues, considering various viewpoints and evidence before reaching informed conclusions.
- The maxi public refers to the general public at large. Panoramic aims to connect the mini public's discussions and recommendations with the maxi public to inform and engage a wider audience.

The divide between the mini public (citizen assemblies) and the maxi public (general population) is a well-known issue in citizen assemblies, as studied by Iten and Moutier in 2022. This divide arises from two main factors. First, while citizens in these assemblies are chosen to be representative, this alone does not make them legitimate, similar

CASE STUDY #1
**MAKE.ORG**

to how polls cannot replace actual voting. Second, as the citizens in the assembly become more knowledgeable about the topic, they develop expertise that may lead them to propose solutions not readily accepted by the general population. This was evident in the case of the French Citizen Assembly on Climate, when their proposal to reduce highway speeds generated backlash, significantly undermining the process (Itten & Mouter 2022).

The initial use, based on a RAG architecture, and described in the previous section, creates a communication channel from the mini public to the maxi public. One future functionality being prepared by Make.org is to ensure debate enrichment through feedback spaces for the maxi public. AI is used here to synthesize the content of testimonials. Citizens are invited to express their views by first showing them this summary and the testimonies of other citizens, before being asked to do so. (Figures 5 and 6)

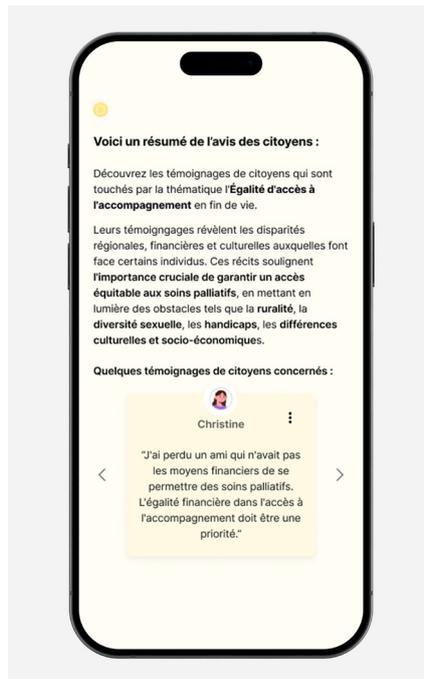

**FIGURE 5.**
This image depicts AI-Generated Testimonial Summaries, and the Citizen Testimonial Carousel on the *Make.org Panoramic platform.*

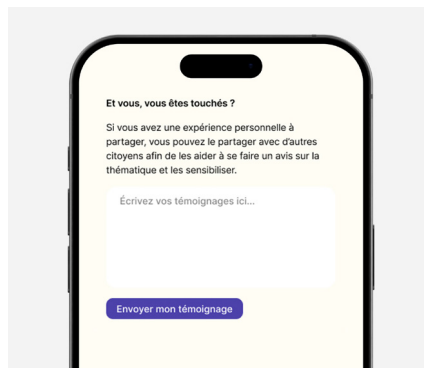

**FIGURE 6.**
This image depicts the Citizen Testimonial Space on the *Make.org Panoramic platform*, as a second step experience.

CASE STUDY #1
**MAKE.ORG**

While AI offers promising opportunities for open prompts and chatbots, these functionalities can in themselves be difficult to grasp for some people. Make.org must support the user in their experience with the interface, in particular by guiding them through the writing process. Here are two very simple illustrations that will be tested in the Panoramic platform (Figures 7 and 8)

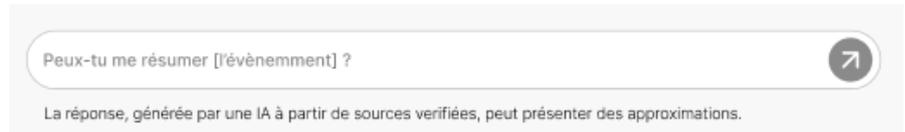

**FIGURE 7.**
This image depicts a placeholder in the open prompt bar to guide citizens on the *Make.org Panoramic platform*.

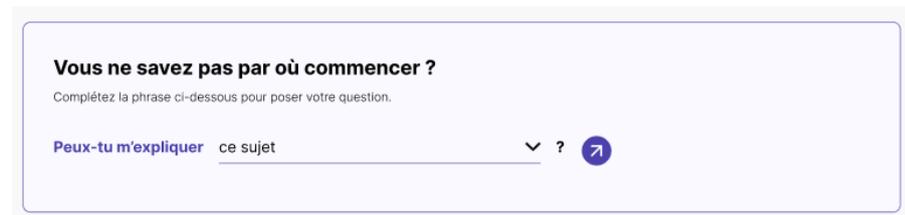

**FIGURE 8.**
This image depicts a drop-down list for triggering prompts on the *Make.org Panoramic platform*.

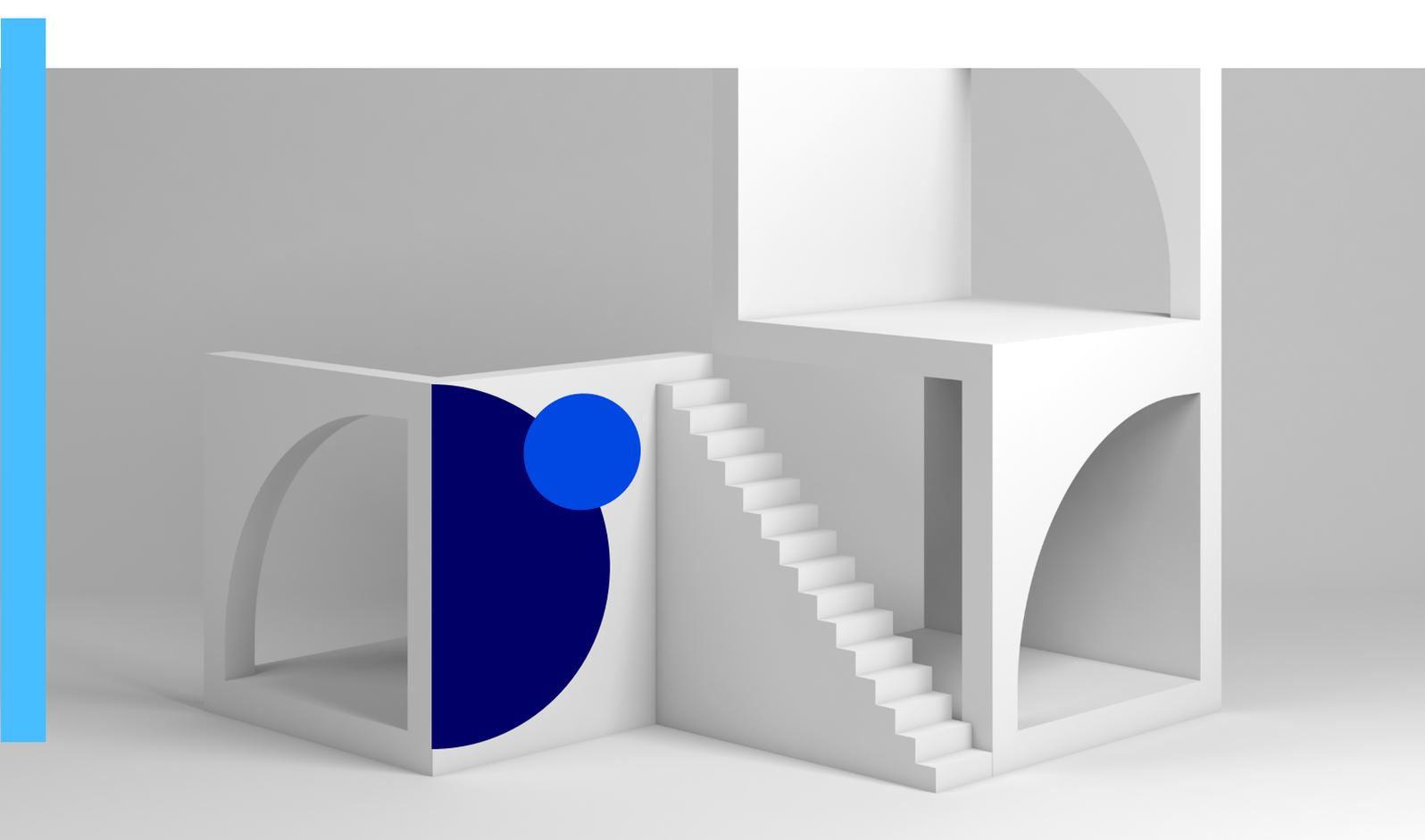

# DIMENSION #3

## SUPPORTING DELIBERATION

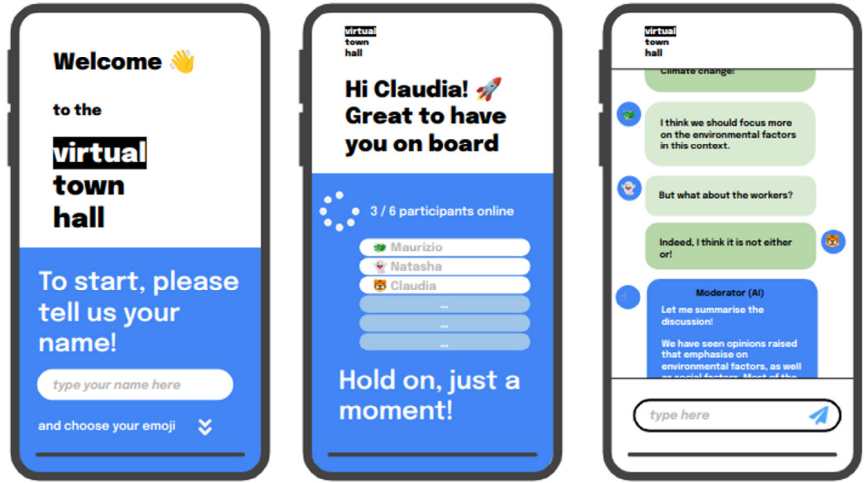

FIG. 12

**FIGURE 9.**
Depicting the screen of the envisioned AI powered online deliberation Platform by *Make.org*.

Make.org views generative AI as a significant opportunity to scale the deliberation process to a larger audience. Deliberation offers a valuable experience for participating citizens (Herzog et al. 2022, Knobloch 2022), enhancing their knowledge and understanding of the issues at hand, while fostering democratic attitudes and skills. However, until recently, deliberation was conducted offline, necessitating the physical presence of all participants and numerous facilitators to guide constructive discussions.

Online participation has since overcome these time and space constraints, enabling Make.org to engage over 100,000 participants in their online consultations. Despite this progress, human facilitators remained essential for effective deliberation.

## AUTOMATING FACILITATION

Even with the use of digital tools, human moderators remain essential. One of the largest deliberations to date was the 21st Century Town Meeting (Lukensmeyer & Brigham 2002), which employed software to synchronize small groups deliberating simultaneously. This approach was deployed in the 2002 "Listening to the City" event, which allowed New York citizens to choose the project for the 9/11 memorial in Lower Manhattan. Over 4,500 diverse participants engaged in structured, small-group discussions facilitated by technology, allowing for both intimate dialogue and large-scale idea sharing. Real-time data collection and synthesis enabled immediate feedback on key issues. This process revealed significant criticisms of the initial redevelopment proposals, ultimately influencing the final design choices for the 9/11 memorial and surrounding area.

However, this system still relied on human facilitators to summarize discussions within each group and help citizens converge toward a group consensus. Recent advancements in AI now enable AI to assume the role of facilitator. AI can effectively summarize discussions, including debates, and the conversational capabilities of generative AI allow it to guide debates toward a constructive conclusion.

Even before the advent of generative AI, Fishkin's team at Stanford managed to design a basic AI facilitator for deliberative polling (Gelauff et al. 2023). With the latest advancements in generative AI, the possibilities have expanded significantly.

Realizing the potential of generative AI for scaling deliberation is a priority for Make.org, which is developing a platform that allows deliberation at scale through a research program in collaboration with Sciences Po and Sorbonne-CNRS. Figure 9 depicts the main screens of such a platform. The first screen is an onboarding screen where users register, followed by a waiting room where they wait for other citizens to connect. Once enough participants are connected, a deliberation room is created, and the AI-assisted deliberation can begin.

The AI facilitator's role is twofold: first, to ensure proper behavior in the deliberation room and moderate disrespectful interactions; second, to advance the discussion by asking questions and providing intermediate summaries to prompt new ideas. At the end of the discussion, the AI will generate a summary, which participants can validate or amend.

Before deploying AI as a facilitator, it is crucial to demonstrate that the AI is unbiased and capable of maintaining neutrality in its role. The AI must effectively steer the debate impartially. Additionally, it is essential for participants to trust the AI facilitator and recognize its positive impact on the deliberation process.

### BUILDING TRUST WITH AI

Even if AI performs well, it still needs to earn human trust. Trust is a primary factor driving the adoption of AI by users (Choung et al., 2022). One of the strengths of Make.org's Panoramic platform is its transparency, providing links to sources used to generate answers. This allows curious users to verify that the AI is accurately summarizing information rather than fabricating responses.

Technology alone is insufficient; it must be tailored to meet citizens' needs and constraints to ensure participation.Make.org seeks to make citizen participation attractive and engaging, which is essential for scaling participation.Make.org has already succeeded in engaging millions of citizens on the Consultation Platform and is leveraging this experience to make engagement with the Panoramic platform even more attractive.

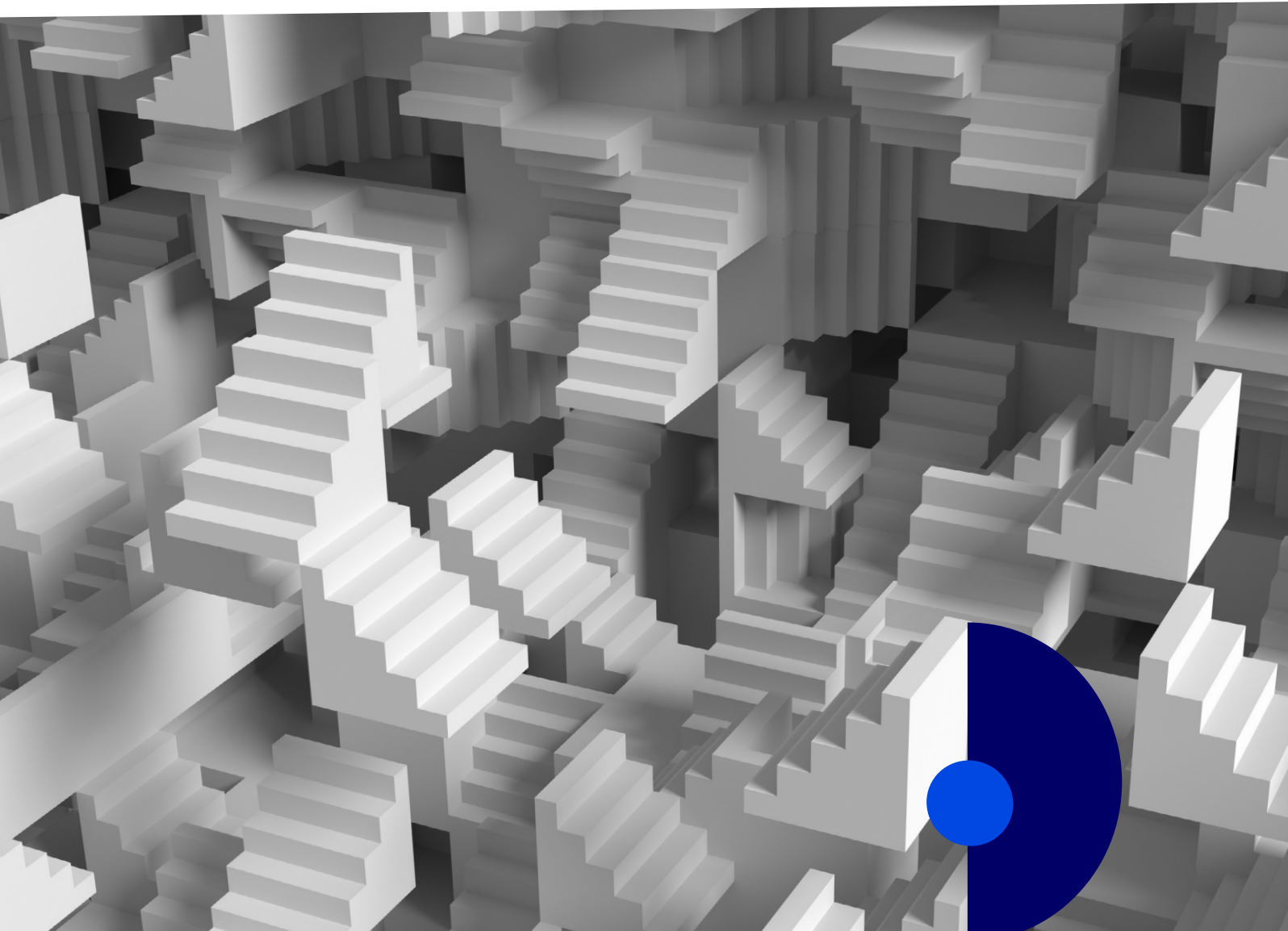

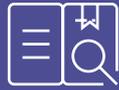

# DIMENSION #4

## SYNTHESIZING INSIGHTS

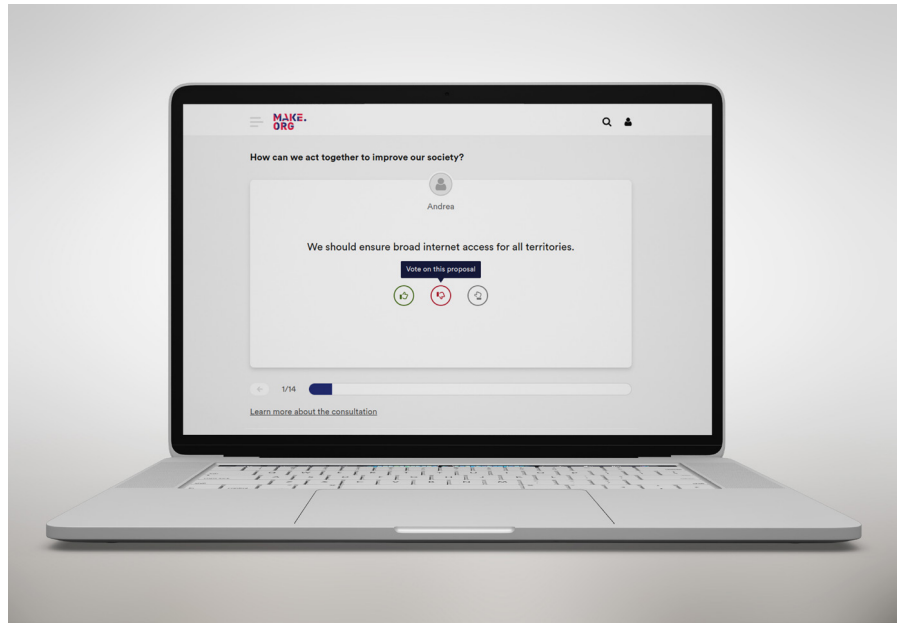

**FIGURE 10.**
Showing a 140-character proposal on the Consultation Platform from Make.org

Make.org is deeply committed to creating meaningful impact. It believes that participation is only valuable if it leads to action, and action can only occur when three conditions are met.

First, the participation process must be well integrated into the decision-making framework, with the institution committed to addressing citizens' priorities. Second, participation must be widespread and balanced in terms of gender, age, and geographical location to ensure the legitimacy of the results. Third, the outcomes of the participatory process must be concise and clear to make them actionable.

Make.org is aware that these three conditions are crucial and designs its platform with all of them in mind.

## INTEGRATION WITH DECISION MAKING

First, the participatory process must be well integrated into the decision-making framework. It is crucial that citizen participation be embedded at various levels of decision-making. This is why Make.org offers multiple platforms, each targeting different phases of the decision process.

At the beginning of a project, when all options are open, wide engagement is needed through the Consultation Platform (Figure 10). By combining a large number of citizen proposals and votes on these proposals, it generates what is called the citizen agenda: a map of the most agreed-upon and controversial proposals. The decision-making process can then be built on this citizen agenda.

If the project is already better defined, with only a few options remaining and only amendments possible, the Dialog platform is more suitable. The project draft can be shared with citizens for their reactions, proposals for improvements, and enrichments. For example, this approach was used for the law on influencers in France, drafted with the Ministry of Economy. The consultation collected 4,800 comments on twelve key measures of the law, helping to improve the draft by adding plastic surgery to the list of prohibited promotional subjects and ensuring expatriates are also bound by the law.

Using the right platform at the right moment is key to making participation impactful.

## MASSIVE PARTICIPATION

Second, for a participatory process to be meaningful and legitimate, it must attract a large and diverse group of participants. Digital platforms are ideal for reaching a broad audience. However, legitimate participation is not solely about large numbers; it's essential to include citizens of varying genders, ages, and geographical locations.

To achieve this diversity, targeted ads on social networks are effective in reaching a wide range of people. The demographics of participants must be monitored to ensure balanced representation. A typical Make. org consultation attracts between 50,000 and 100,000 participants, with balanced representation in terms of gender, age, and geographical location. When balanced participation cannot be ensured, statistical tools are employed to reweight the votes, making the results representative.

## ACTIONABLE INSIGHTS

Third, it is crucial to ensure that input from participants is synthesized into meaningful and impactful results. Prioritizing valuable and actionable input over sheer volume of data is essential. That's why Make.org's Consultation Platform, designed to collect thousands of proposals and votes, limits citizen proposals to 140 characters and does not allow comments. When there are many proposals to vote on, short and meaningful suggestions are more effective than lengthy texts. Comments on thousands of proposals tend to be too dispersed to be useful.

In contrast, the Dialog platform, dedicated to gathering feedback on a few well-defined projects or recommendations, relies on comments as the primary mode of expression. This approach works because comments are aggregated around a common subject and are usually constructive.

This is the "I need you" approach: only request contributions from citizens if they are genuinely needed and can be utilized. This principle respects citizens' time by avoiding unnecessary tasks and focuses on collecting valuable and usable data. The data must then be distilled into meaningful insights through a robust synthesis process.

In conclusion, respecting users' time by avoiding unnecessary tasks, ensuring efficiency in managing input by focusing on valuable data, and maintaining confidence in the synthesis process are all essential for effective participation.

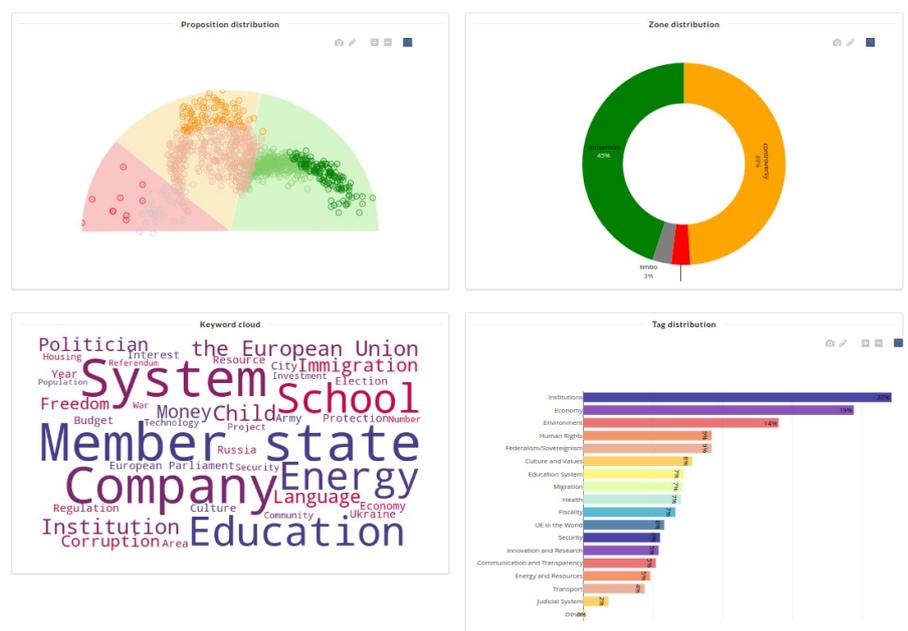

**FIGURE 11.**
Screenshot of the internal AI analysis tool for the Consultation Platform at Make.org

CASE STUDY #1
**MAKE.ORG**

### AI FOR SYNTHESIZING INSIGHTS

Once these three conditions are met, AI plays a crucial role in synthesizing debates and transforming citizen contributions into meaningful insights. Make.org has been leveraging AI from the beginning to analyze large consultations, categorize proposals into topics, and group them into coherent ideas. However, AI has always been used under human supervision. The human-in-the-loop approach, where humans provide the analytical framework, instruct the AI, and validate or correct its output, has been the most accurate method until recently. Figure 11 shows Make.org's internal analysis platform used to oversee the analysis of a Consultation.

With the advent of generative AI, the days when analyzing citizen participation required a data scientist are gone. Generative AI now empowers individuals with AI-driven insights. While validating outputs and training effective AI models still require expertise, advancements in the field have made powerful analysis tools accessible even to small-town teams with a single dedicated citizen participation representative. Make.org, along with many others, is now developing tools to enable non-technical users to leverage AI for analyzing citizen inputs. With the advent of ChatGPT, gone are the days when analyzing citizen participation required a data scientist. Generative AI now empowers individuals with AI-driven insights. While validating outputs and training effective AI models still require expertise, advancements in the field have made powerful analysis tools accessible even to small-town teams with a single dedicated citizen participation representative. Make.org as well as many others are now building tools to enable non-technical people to leverage AI to analyze the citizen inputs.

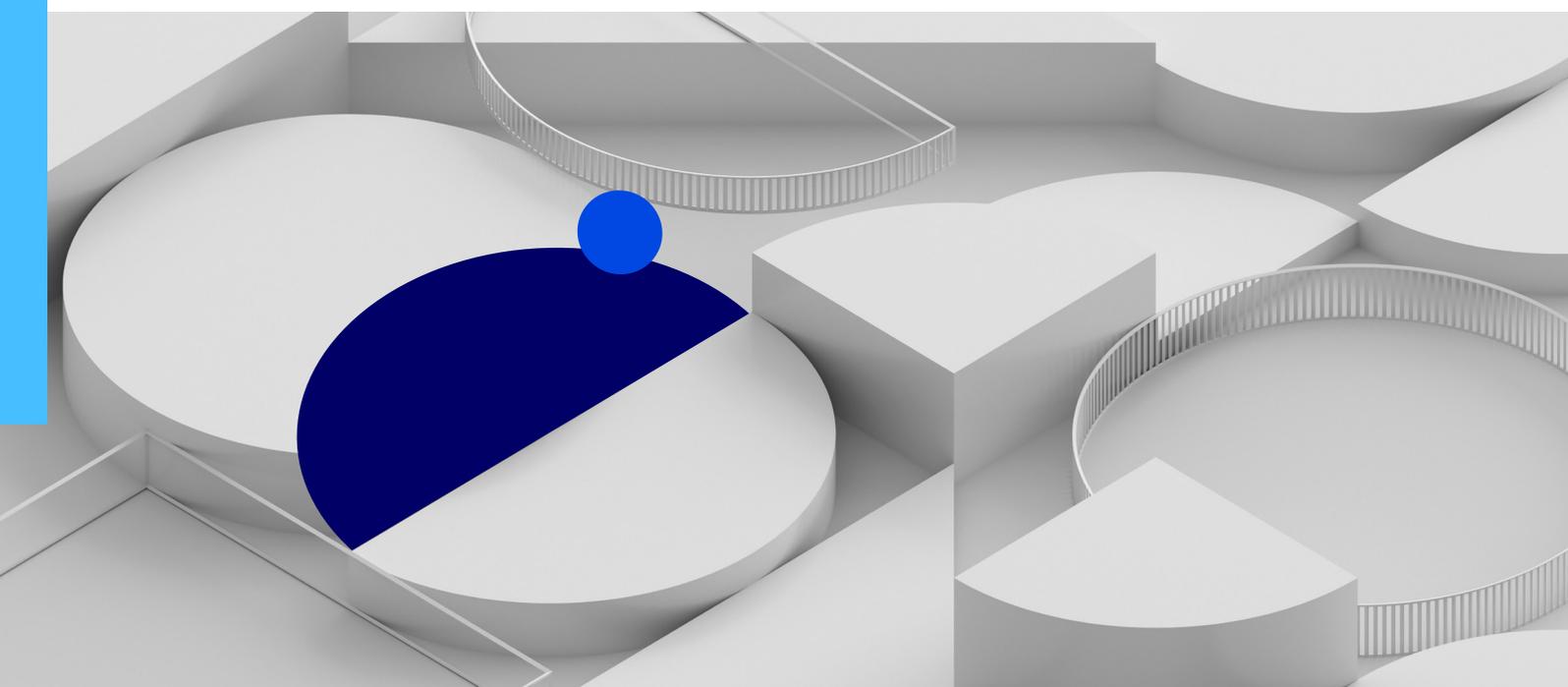

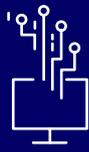

# DIMENSION #1

## MAKING INFORMATION ACCESSIBLE

A primary feature of MAPLE is making information about the content and process of pending legislation more accessible, and we have several AI integrations in development to enhance that use case: bill summaries, bill tagging, and redlining.

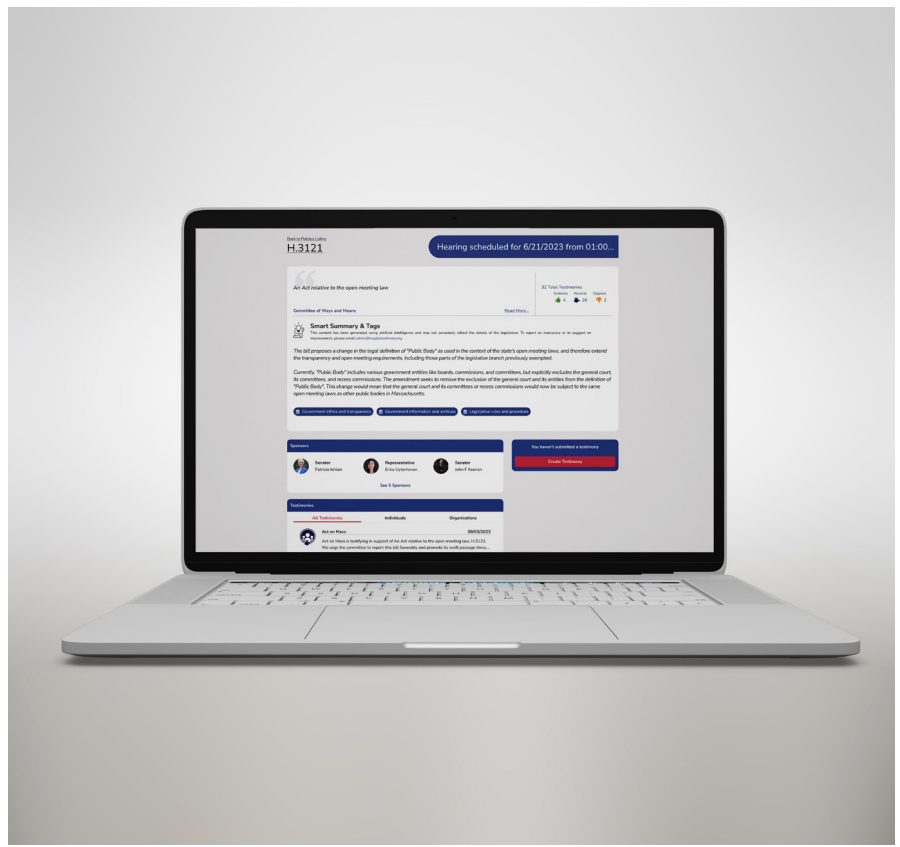

**CASE STUDY #2
MAPLE**

**FIGURE 12.**
Screenshot of the in-development feature of the MAPLE platform showing the AI-generated summary and tags for bill H.3121 in the 193rd General Court (MA legislative session).

CASE STUDY #2
**MAPLE**

## BILL SUMMARIES

MAPLE is developing Large Language Model (LLM)–derived summaries of legislation to be published alongside every bill to help users understand the purpose and impact of each bill. These brief summaries (e.g., 1 to 3 paragraphs) are being generated at multiple reading levels and will be displayed alongside bill text. The OpenAI GPT-4o model is used for this work.

Because bills can be complex, lengthy, heavily interlinked, and written in highly domain-specific terms, they are challenging targets for AI summarization. To improve reliability, as much context as possible is incorporated into the LLM prompt, to maximize the potential for in-context learning (Dong et al., 2023, Geng et al., 2024). The summarization prompt includes the bill text, referenced sections of the Massachusetts General Laws and their titles, and information about the committee considering each bill. Because legislative drafting is generally done by insertions and deletions to existing law, the context of existing law text is often strictly necessary to understand the function of a bill.

Nonetheless, the model does make mistakes and sometimes incorrectly represents the function of a bill. It is difficult to establish domain-specific performance metrics for abstractive summarization. Human evaluation remains the gold standard, and top performing automated approaches such as ROUGE-L involve first generating human-authored ground truth summaries at scale to measure against reference summaries (Zhang et al. 2024). While MA legislation lacks ground truth, human authored summaries, manual review of generated summaries is carried out to understand performance. Non-monotonicity in legislative language, such as one or multiple negations, presents a particular problem (Han et al. 2024). One sample bill proposes to delete one entity from an enumerated list of exempted bodies established in current public meeting law. By removing the exemption, the bill would effectively extend the existing law to apply to the currently-exempt entity. Even after iterations to improve the prompt engineering, the LLM would frequently summarize the bill's function incorrectly, interpreting that the bill would add, rather than eliminate, an exemption, and would state the bill's likely impact in a way exactly contrary to its actual language.

Bills drafted to affect a variety of issues, including annual budget bills and so-called "omnibus bills," present further challenges for the LLM model. Bills with a large breadth of topics require the LLM model to understand how different bill sections relate or don't relate to each

CASE STUDY #2
**MAPLE**

other, as well as identify and communicate the relative importance of discrete sections. Moreover, the total size of the bill and law sections it references can exceed the context length of the LLM or be costly to ingest (large number of input tokens), requiring the use of RAG. In this situation, LLMs are known to have performance limitations (Liu et al., 2024) that can lead to missing significant details buried within long contexts.

## TAGGING LEGISLATIVE PROPOSALS WITH AI

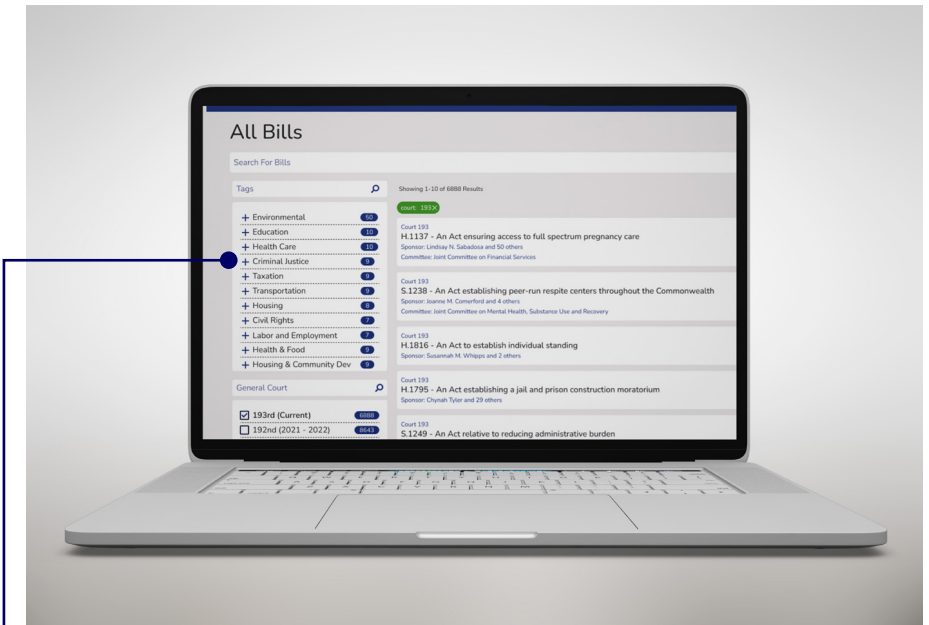

**FIGURE 13.**
Screenshot of in-development MAPLE feature allowing user to filter bills by their AI-assigned topic.

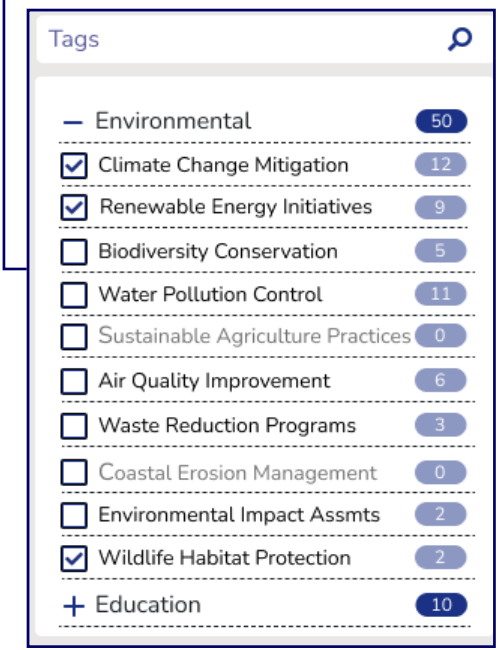

**FIGURE 14.**
Screenshot of expanded view of "Environmental" topic allowing user to select and filter by specific AI-assigned tags.

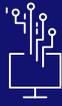

Constituents of Massachusetts are, arguably, overwhelmed with legislative proposals. Massachusetts' legislature files more bills than any state except New York (FiscalNote 2021), more than 6,000 per year on average (Bucchianeri, Volden, and Wiseman, 2024). The only official categorization of these bills is their assignment to committees organized around broad subjects such as election laws or education. Committees are assigned hundreds of bills per session, offering little scaffolding to constituents looking to navigate this mass of legislation.

To aid constituents looking to find filed legislation relevant to their interests, MAPLE is using AI to automatically assign content tags to bills. An LLM labels each bill with tags from a predefined list of 250+ topics, a refined list initially sourced from the Congressional Legislative Subject Terms taxonomy,[10] with adjustments for the state-specific application. A state-specific taxonomy could also be developed manually or by clustering the bills before the MA legislature, but MAPLE expects relatively little variation in taxonomy terms across jurisdictions and its platform is designed to be portable to other states, so MAPLE seeks to align to national taxonomies where possible. The LLM is deployed in a hierarchical multi-label classification setup, first assigning bills to categories and then assigning between one and six category-specific tags per bill. These tags will allow us to expose search filters in the web interface to help constituents easily find bills related to a specific topic.

### BILL COMPARISONS AND REDLINING

MAPLE is further building tools that use LLMs to compare and contrast similar legislation. This feature would help constituents understand diverse legislative proposals to address similar issues and would help them choose between alternative policy prescriptions when generating testimony in favor or opposed to legislation.

Where multiple bills are introduced in a state legislature all seeking to address the same issue, AI-driven comparisons and synthesized analysis could allow a constituent to quickly understand their differences in approach and effect. AI could also be deployed across jurisdictions to identify how other states propose dealing with that issue and to source new ideas.

Lastly, MAPLE is developing features to help constituents understand the specific legal effect of legislation by compiling the bill text into redlined versions of the relevant Massachusetts General Laws. As with some other jurisdictions, MA bills are a series of insertions and deletions into existing law, which makes it difficult to understand the new content without seeing the affected law text and understanding the broader context. By automatically compiling bill text into inline, red-lined edits of the existing law, MAPLE can make legislative proposals much more contextualized and explicit.

---

10   https://www.congress.gov/help/field-values/legislative-subject-terms

In Massachusetts and many other jurisdictions, legislative procedures and norms have been established over several centuries and represent a barrier to entry for new entrants to policy debates. Even stakeholders who have clear and material interests in policy may struggle to inject their perspective into the policymaking process, and legislators sincerely interested in the opinions of and impacts on underrepresented populations may struggle to access their feedback. Many people turn to social media as a familiar and accessible outlet for their political opinions, but material posted there is a "needle in a haystack" and is less actionable to legislators without formal entry in the legislative process.

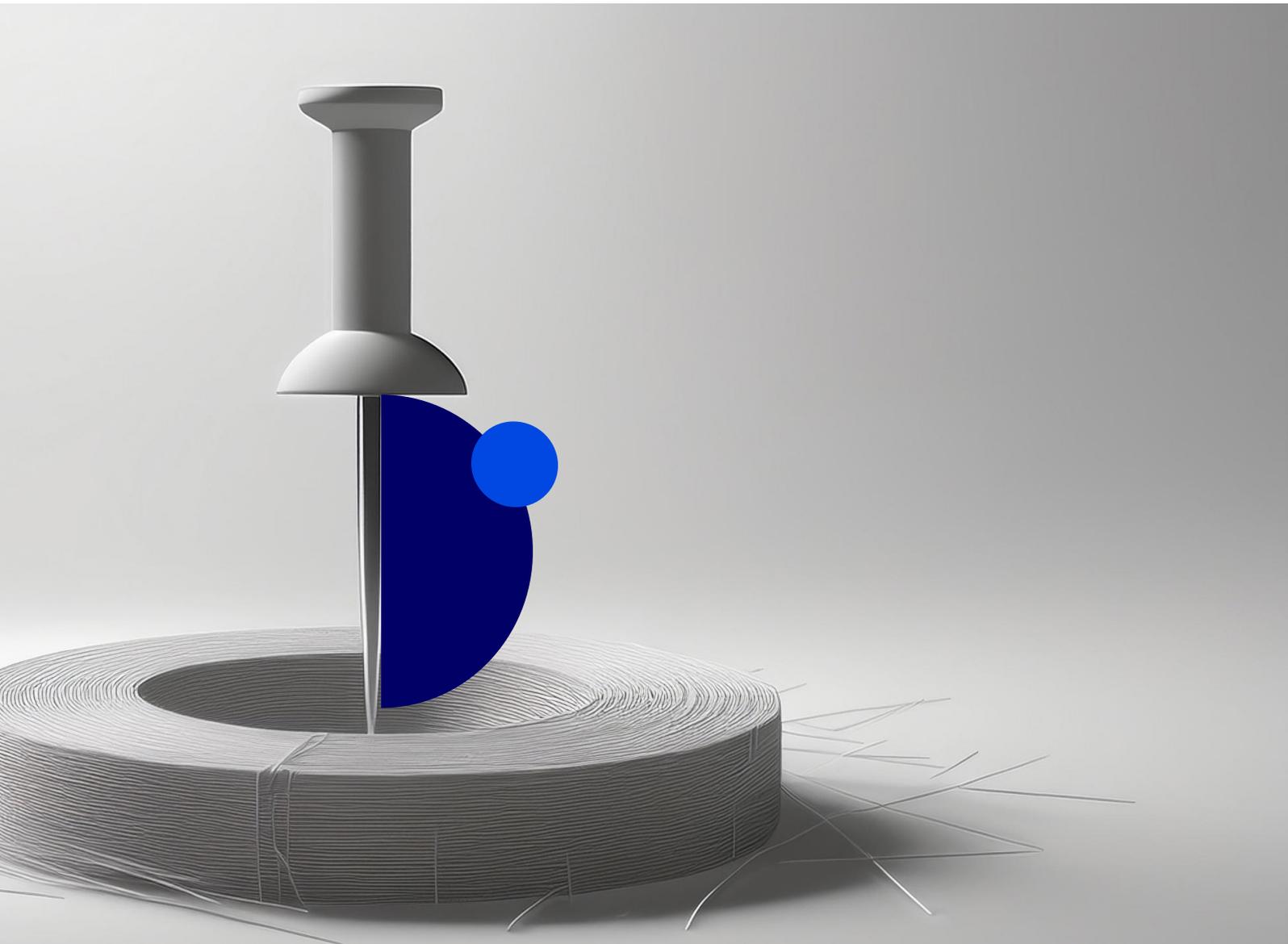

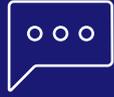

# DIMENSION #2

—————

## FACILITATING EXPRESSION

MAPLE is developing AI features to assist users with drafting legislative testimony and to localize political opinions formed around national issues to the state context.

### DRAFTING TESTIMONY WITH AI

Anyone capable of expressing their viewpoint on an issue in any level of detail in any language should be able to draft appropriate testimony submittable to the legislature. A structured question and response model may be deployed to connect a constituent's broad policy perspective on an issue to the various relevant legislative proposals, to help the user understand the differences in approach and underlying values across the legislative proposals and to select one that best aligns with the user's perspective, and finally to help the user draft relevant and well-formatted testimony on that legislation that articulates their position and any specific adjustments to the legislation that user would like to see.

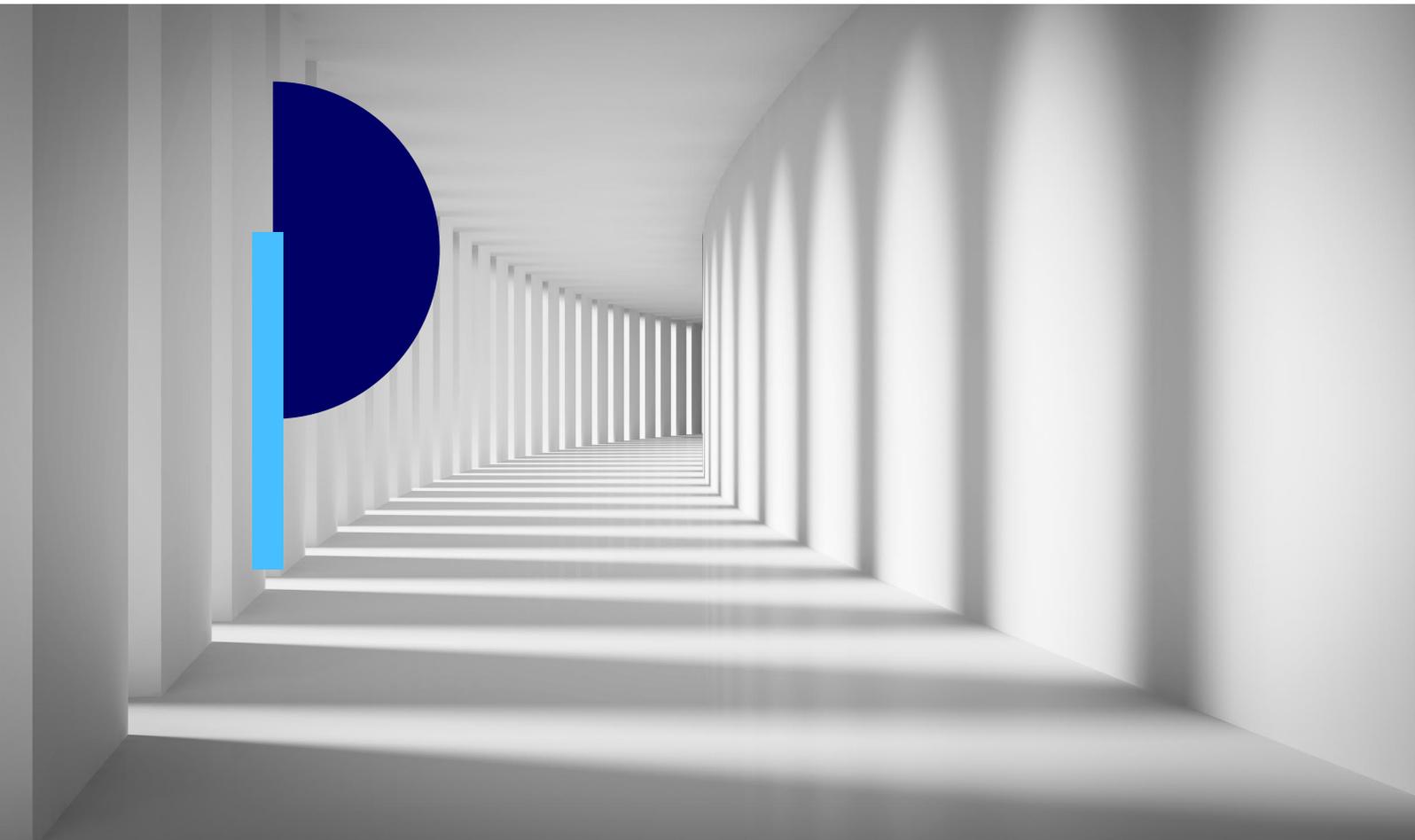

**CASE STUDY #2 MAPLE**

AI can specifically accommodate users in articulating their views with respect to:

1. **References and citations:** Referencing specific bills and sections, committees, and members and citing substantive sources of information to support arguments made in the testimony. These can be achieved by augmenting LLMs with RAG to query from defined databases of existing law and authoritative news or research sources, respectively.

2. **Formatting:** Formatting testimony and adhering to the traditional decorum of the legislature. A chatbot interface can elicit substantive information from the user on their policy perspective and output a properly formatted testimony document.

3. **Translation:** Translating from the user's preferred language to the primary language of the legislature. Neural machine translation (NMT) models can help democracies be more multilingual (Chartier-Brun & Mahler 2018).

4. **Soundness:** Helping users refine their position, considering its positive and adverse consequences. A chatbot interface can provide iterative feedback to a testimony author to sharpen their expression.

## HELPING CONSTITUENTS CHANNEL EXISTING PREFERENCES ON LOCAL LEVEL

Another proposed feature would enhance civic engagement by bridging the gap between national political affinities and local legislative actions. Many members of the public have strong alliances with prominent national politicians and movements, but often these preferences are not expressed at the state and local levels. This feature would solicit and connect users' existing perspectives to local representatives and pending legislation. By tying constituent perspectives to their local environments, this feature allows people to contribute more meaningfully in local and state politics. The tool would provide an overview of legislative alignment, highlight gaps, and informs users about ongoing legislative efforts.

When a user inputs a notable political figure—for example, a US senator—the feature would output the broad policy changes required to align the jurisdiction with that political figure's platform. For instance, it might identify the need for more publicly funded housing, universal healthcare, or enhanced workers' rights. The tool would then identify and present the state legislative proposals best aligned with achieving these high-level changes.

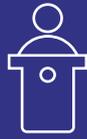

## DIMENSION #3

### SUPPORTING DELIBERATION

MAPLE's current focus is not on facilitating direct conversation between users, but rather in supporting the broader conversation between the public and the legislature. MAPLE's AI integrations in development will help synthesize the body of testimony submitted by users and to help find consensus within that corpus.

### AN AI-ENHANCED REPOSITORY OF TESTIMONY

In addition to searching, tagging, and annotating bills, similar tools can be deployed to enhance understanding of testimony received from constituents. For bills that have received heavy advocacy and many submitted testimonies, reading them individually will be beyond the capacity of a typical constituent or legislative office.

LLM-powered testimony summarization helps us to synthesize the perspectives from diverse constituents. The MAPLE team has developed both abstractive and extractive prompts (Pilault et al. 2020; Schneier and Sanders 2023).

- Abstractive prompts, such as queries to provide a brief summary of all received testimony or to compare and contrast the values cited by individuals on each side of advocacy for a bill, provide brief synopses of all the submitted testimony.
- Extractive queries, intended to highlight examples of testimony from public officials or the most cogent arguments made in testimony, return specific testimony from the broader corpus.

Through this development, RAG has worked well for both the abstractive and extractive testimony summarization tasks. In particular, this system provides a natural way for the LLM output to reliably cite sources, in terms of specific documents (testimonies) that were considered in generating the output. These explicit citations can be highly useful in constraining the LLM to rely on factual information in its output (avoid hallucinations) and to build confidence among users that the results are verifiable and accurate.

CASE STUDY #2
**MAPLE**

A full implementation of this system would give the user some control over the query to execute across the testimony. Rather than pre-populating a handful of pre-generated queries, the user could formulate and receive responses to their own questions about the testimony, on demand. This could provide a valuable tool for journalists who may seek to quickly understand public sentiment on a range of bills or to identify individuals for further conversation. MAPLE has also worked with stakeholder groups in the public health space to define and refine queries of constituent testimony useful for their own advocacy work, such as,

**1.** Queries for testimony delivered by specific constituent groups, such as public officials, frontline health workers, youths, or parents.

**2.** Enumerations of values cited by groups testifying from either end of the spectrum on polarized issues.

**3.** Enumerations of lines of evidence cited in testimony across the political spectrum.

**4.** Examples of highly personal testimony that relates lived experience of the testifier.

**5.** Examples of misinformation propagated by testimony.

### TESTIMONY CURATION AND CONSENSUS FINDING

A related feature can make it much more efficient for a user to understand the arguments and values discussed within the testimony of a given bill, while also incentivizing future users to contribute more thoughtful testimony that addresses the range of expressed perspectives and arguments. While social media deploys algorithms that highlight and incentivize outrageous and polarizing content, AI can be deployed to highlight representative user submissions and incentivize meaningful deliberation. This approach curates testimony according to how effectively it represents the range of perspectives within existing user submissions and how well it articulates and responds to the range of value-based arguments that are expressed within that corpus of public testimony. Elevating testimony on this basis incentivizes submitters to address key arguments and counterarguments properly. By highlighting well-rounded and responsive testimonies, it encourages individuals to engage deeply with the topic, consider opposing viewpoints, and contribute constructively to the discussion. This improves the quality of testimonies and enriches the overall discourse.

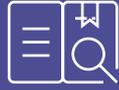

# DIMENSION #4

___

## SYNTHESIZING INSIGHTS

### SUBMITTING TESTIMONY

MAPLE addresses this gap by making it easier for all communities to contribute feedback through established channels in the legislature. MAPLE users who submit testimony on the site are encouraged to submit that testimony to the relevant legislative committee with just one additional click. The implementation of this feature launches the users' own email client to send their testimony from their own email account, cc'ing the committee chairs as well as their own representatives, establishing direct connections between legislators and their constituents without intermediation from MAPLE. Furthermore, the email is formatted to address the committee chairs as the recipient and to directly reference the appropriate bill number.

This feature makes it easy for users to submit testimony through official processes, as both the MA House and Senate rules provide for submission of written testimony to legislative committees by email. Furthermore, by increasing the accessibility of this process, MAPLE advances the stated intention of both chambers to increase the diversity of people submitting testimony (Massachusetts General Court, n.d.-a, Section 17e; Massachusetts General Court, n.d.-b, Section 12).

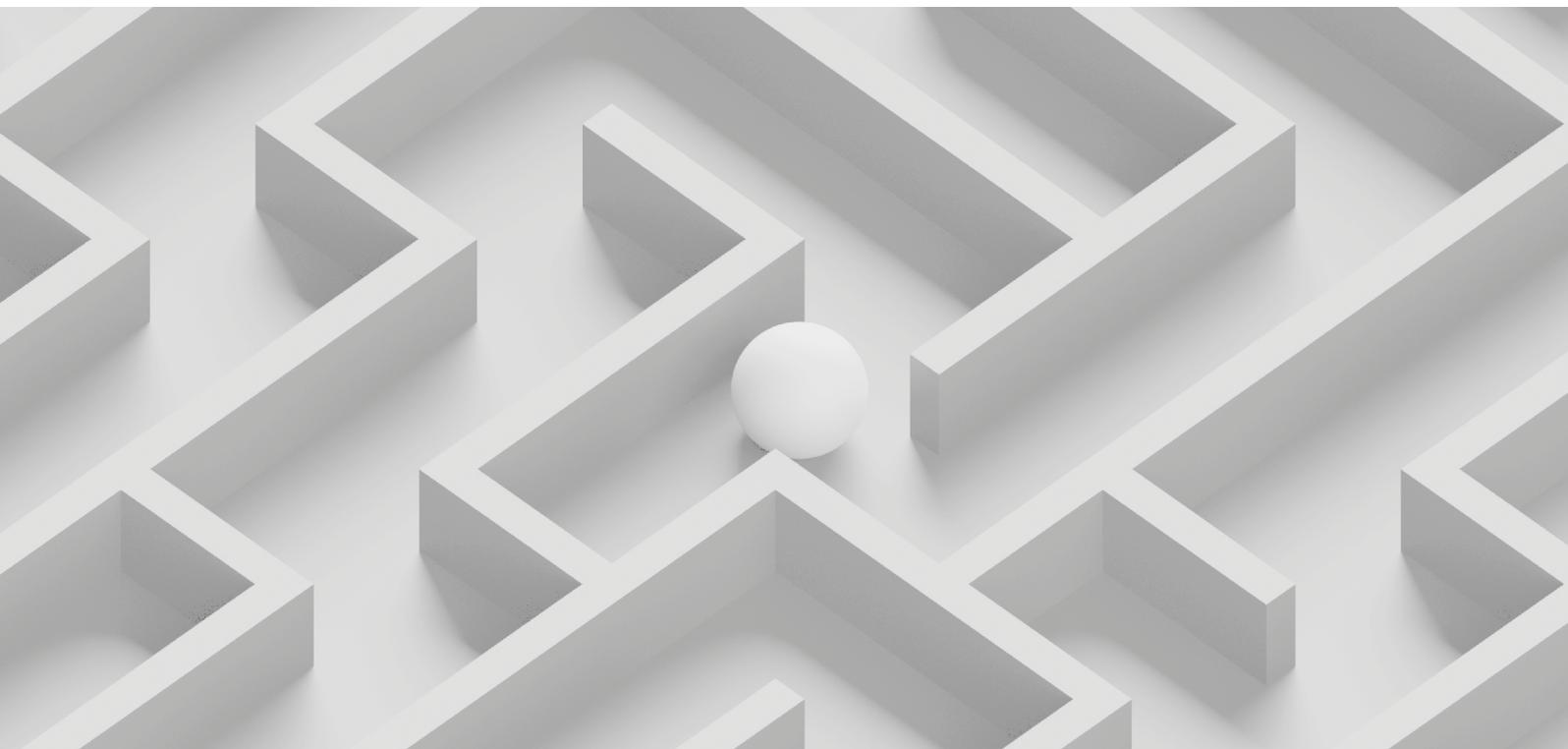

CASE STUDY #2
**MAPLE**

### AI SYNTHESIS FEATURES

AI integrations can help the legislature incorporate large volumes of written testimony into their deliberations in actionable ways. Automatic summarization of testimony can help committees identify points of consensus and disparity across public testimony. AI could be used to automatically cluster testimony by stakeholder communities, for example, to highlight testimony emerging from environmental organizations versus businesses and commercial associations. A feature that automatically compares testimony received to testimony on similar past bills and to other forums of public debate, for example, through social media and web queries, can help expose stakeholder communities that have not yet engaged in the legislative process and could be brought into the fold through outreach. These features can inform legislators on how to adjust legislation to reflect established consensus or to damper the negative consequences on a specific stakeholder group that opposes such consensus.

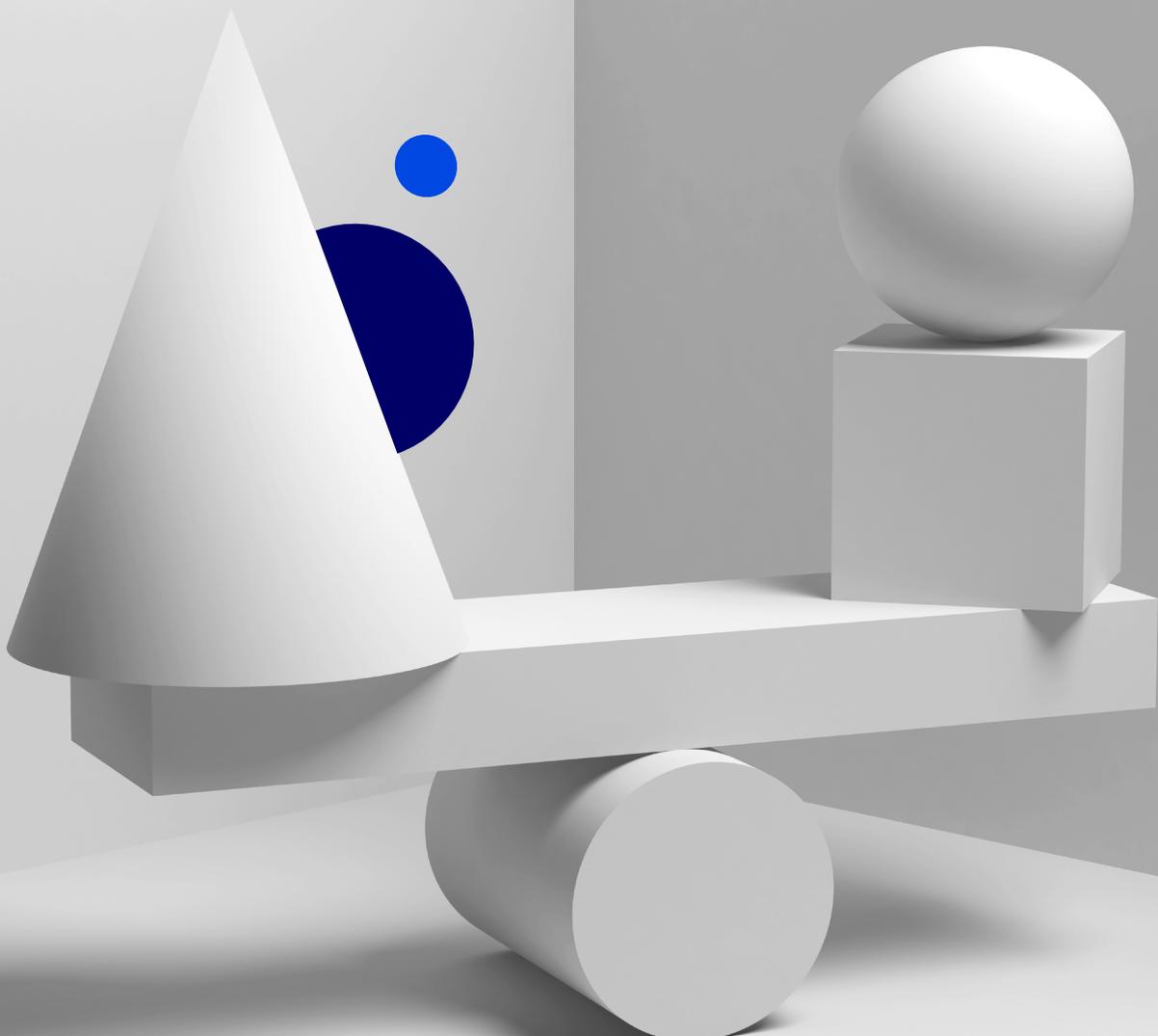

# DISCUSSION

In this section, we synthesize the philosophy and experiments of the Make.org and MAPLE teams in terms of key comparisons and contrasts, including with respect to jurisdictional differences. We then discuss the risks of AI applications to legislative engagement and conclude with recommendations for both governments and civic organizations.

## ▌ COMPARISON AND CONTRASTS

Both Make.org and MAPLE seek to address fundamental threats to liberal democracies: increasing polarization of opinions, growing distrust towards elected officials and a political system's deteriorating ability to effect change and improve citizens' well-being. Although Make.org and MAPLE operate in different jurisdictions and have adopted different approaches, their goals and principles are aligned. Make.org focuses on facilitating consensus among a large number of constituents on various topics, while MAPLE integrates discussions and consensus-building efforts into the legislative process. MAPLE operates within the two-year legislative session in Massachusetts, whereas Make.org's tools can be utilized for a broader range of deliberative projects and scopes. Both platforms seek to leverage advanced technology to reconnect citizens to one another and with elected officials, enabling them to participate meaningfully in political and collective decision-making.

The different types of deployments for Make.org and MAPLE likely require different metrics to evaluate their success. Beyond the obvious measurements of engagement that both employ (i.e., clicks and active daily users), Make.org deployments are often initiated for a specific purpose (e.g., finding consensus on a specific policy issue) and success can be evaluated by the soundness of consensus reached and how representativeness of the participants. MAPLE, its operations tied to the legislative sessions, may be better evaluated by comparing traditional democratic health indicators (e.g., electoral competitiveness, number of bills passed, etc.) across legislative sessions. The shorter turnaround on Make.org metrics may better facilitate iterative improvements to their tools; however the latent MAPLE metrics would more directly indicate whether their platform is improving democratic functioning.

These projects may also complement each other. For example, platforms like Make.org may pull content from platforms like MAPLE to inform the content and prompts of a deliberation project, and then feed the outputs of that project back into a platform such as MAPLE with a community ready to mobilize around specific legislative changes.

Another difference lies in the business models of our organizations. MAPLE is a non-profit organization that relies on charitable donations to operate. This public support can be advantageous for advancing the institutionalization of the platform but can also limit the available funding for its development. While funding could limit the MAPLE organization from expanding, as an open source webtool, other initiatives can clone the MAPLE codebase and deploy it in their own jurisdictions.

In contrast, Make.org chose to be an independent, mission-driven for-profit company. Make.org is paid to deploy and operate its platform for institutions and companies, and has developed products designed for different scales of participation. This business model allows Make. org to finance the development of its platform but may limit its ability to become institutionalized.

It is too early to determine which path provides the best option for both sufficient funding and maximum institutionalization of the platform, and it may differ by context. However, it is worth noting that there are multiple possible approaches. One certainty remains: it's by combining our efforts and learning from one another that we will increase the likelihood that these platforms are adopted by local and national governments.

Regarding jurisdictions, the US political environment is highly nationalized and polarized. These, among other factors, make it difficult for citizens to have influence at the federal level. For state-level participation, the decline of local news and attention to state politics makes it difficult for individuals to find an entry point to engage with local lawmaking processes. State legislatures present a natural point of leverage to deploy impactful but well-tailored digital infrastructure. The EU is also grappling with increasing polarization of opinions and growing distrust towards politics and democracy. However, the political environment in the EU is more conducive to citizen participation. Many initiatives are taken at the regional, national, or European level, like the deliberative commissions of the Brussels

Parliament, the "Grand Débat National"[11] and the Citizen Convention on Climate Change in France, or the Conference on the Future of Europe. In that context, Make.org uses digital technology to substantially widen the scope of public deliberation at the regional, federal, or multinational level, informing the agenda for policymakers across multiple levels and jurisdictions.

Another important jurisdictional consideration is the public's familiarity with digital deliberative tools. Citizen assemblies are more common in Europe than in the US (Reuchamps et al. 2023), and thus European civic projects face fewer educational and go-to-market barriers than their US counterparts. Europe also has significantly more advanced adoption of regulation of AI technology. See, for example, the EU AI Act, which may carry both benefits for public trust and acceptance of the technology, as well as awareness and skepticism of its inherent risks. Outside of the US and Europe, developing economies and governments with fewer resources may have difficulty adopting AI technologies.

**Despite these jurisdictional differences, it is striking that both MAPLE and Make.org have developed tools to connect citizens with legislative bodies. They both leverage AI to facilitate citizen feedback and participation in the lawmaking process. These inter-continental similarities suggest that they are some core common building blocks for participative technologies.**

11  Both the "Grand Débat National" and the Citizen Convention on Climate Change were responses to the Yellow Vest movement in France. The Grand Débat was a participation success, with over 1.5 million participants engaging online and in local meetings. Additionally, more than 500,000 written contributions were submitted in the "Cahier de Doléances." However, synthesizing this large and diverse corpus of contributions proved challenging. The synthesis was incomplete when officially presented, and the "Cahier de Doléances" remains unavailable for researchers. The Grand Débat demonstrated that large public consultations can be ineffective if their impact is not well planned in advance. Consequently, it has not been replicated.

## ▌ RISKS AND MITIGATIONS

The risks we face are endemic to other public-facing integrations of AI technologies. We must overcome potential mistrust of our use of AI and backlash if any AI-generated content is found to be false or perceived as biased or misleading. While MAPLE and Make.org are both keenly aware of and seeking to mitigate any bias within our models, the potential for failure will remain.

In order to mitigate these risks, Make.org has launched a research project called Democratic Commons to define, identify, evaluate, and correct the democratic bias of the LLM. This project aims to deliver a public tool to evaluate LLM bias that can undermine the democratic uses of AI and provide open-source corrected LLMs safe to use in a democratic context.

Identifying and mitigating the bias of the LLM is a critical task. Once an LLM starts doing tedious tasks for humans, like summarizing long texts, humans tend to rely on them blindly. LLM biases are then a very insidious risk. Let's assume that an LLM tends to favor certain ideas because they appear more frequently in its training set. Those ideas will appear more often in the summaries generated by AI, or in a more prominent place. In the end, the training of the LLM could decide the outcome of the deliberations rather than the actual arguments exchanged.

Once the biases are managed, we still need to make sure that AI will be able to steer the debates towards a constructive outcome. This task is more a matter of prompt engineering and testing, but it remains to be proven that AI can perform well for these critical tasks. Make.org and MAPLE share two key principles for responsible AI deployment. First, we believe AI should be used to reduce, not exacerbate, existing inequalities. Second, we emphasize the importance of building public trust through transparency and responsiveness to feedback. We mitigate these risks through clear and specific disclosure of uses of AI, through limiting AI integrations to use cases with limited potential for harm, and through testing of performance against real-world data.

An ever-present risk in digital processes is hacking through the use of bots or other trolling techniques. To combat this, Make.org employs both technical and data solutions. The Consultation Platform uses captchas, anti-ddos systems, and cryptographic voting keys. On the data side, fraud detection algorithms help identify and counter attacks.

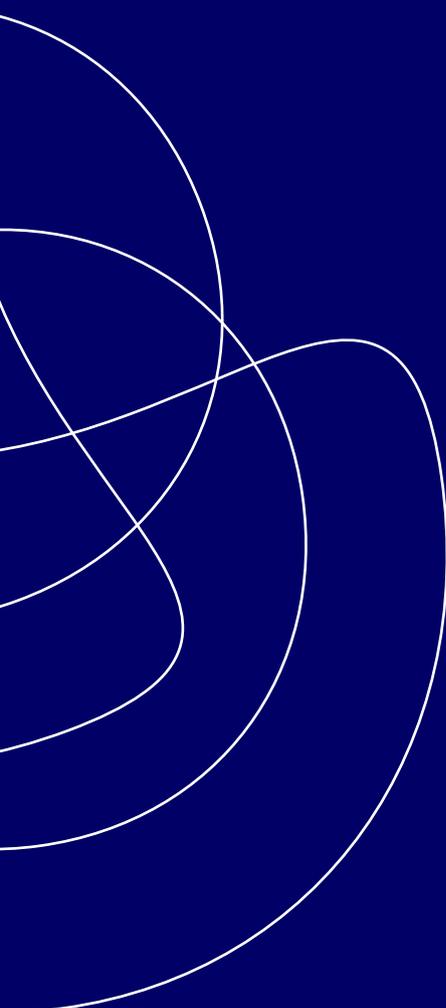

MAPLE mitigates the harms of "trolls" and other bad actors by validating the identity of organizational users against corporate filing records. This ensures that organizational testimony, which is emphasized throughout the website, is genuine. MAPLE also limits opportunities for reductive actions by not allowing comments on testimony and restricting users to one submitted testimony per bill.

Relatedly, we operate our projects against the background of global social concern over political apathy (Zhelnina 2020). Our projects are aimed at increasing participation, but manifesting success will require attracting broad, diverse, and sustained engagement in spite of these trends.

## ▌ RECOMMENDATIONS FOR GOVERNMENTS AND CIVIC ORGANIZATIONS

Our recommendations for governments and legislators exploring AI-enhanced deliberative initiatives are grounded in the goals of access, trust, and impact.

We recognize that civil society cannot sustainably solve these problems while acting alone. Both our projects seek to demonstrate, within a civil society context, how AI can assist legislative policymaking. Our hope is that institutions of government will recognize the value of these capabilities and integrate them into their institutions over time, while organizations like ours continue to innovate further new techniques. Interestingly, local (sub-national) jurisdictions may often be best positioned to adopt these technologies at the governmental level, since they can do so with more agility than larger governments. However, civic organizations may be best positioned to operate these technologies at a larger / national scale, since it can be difficult to attract funding to tackle the problem in more local contexts.

In order to attract institutional adoption, civil society organizations piloting platforms such as these must demonstrate value. Early adopters among policymakers will be attracted by features targeting legislators, such as convenient access to input from their own constituents, analytics about their own bills and policy proposals, and easy to use engagement tools that they can recommend to their constituents. Once a critical mass of early adopters is reached, those legislators' colleagues will be more likely to use and recommend the platform, and thus may be more likely to adopt its features within their own institutions' websites.

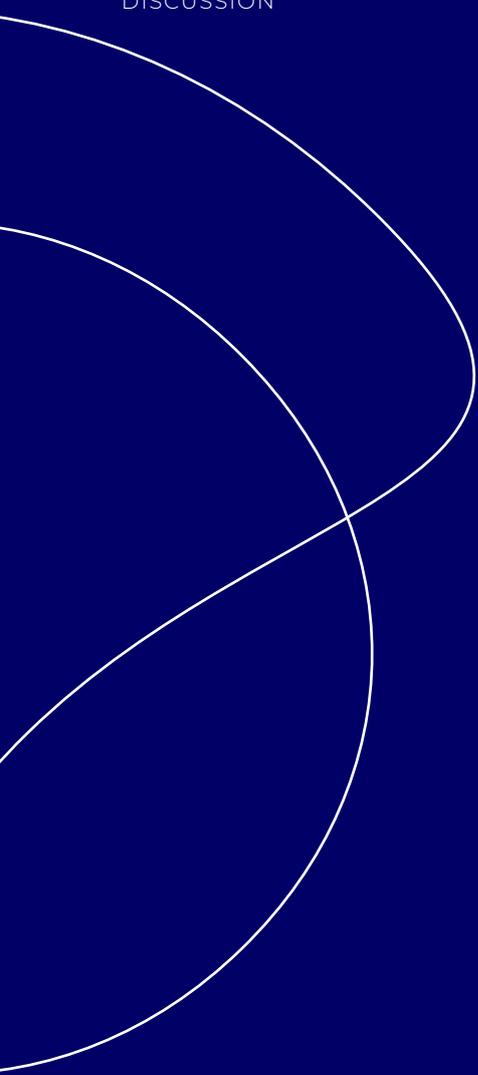

Deliberative projects must be accessible across several different criteria, considering cost, convenience, language translation, and cultural norms. Broad participation is important to ensure the inputs and outputs are diverse and representative of community stakeholders. By increasing access and conveying the accessibility of the platform clearly, projects are likely to increase buy-in and trust in the platform. The 2021 Global Assembly on the climate and ecology crisis exemplified these values, by carefully planning a transparent participant-selection process that valued diversity and inclusion, and compensating all participants equally.[12]

Trust should be a foremost consideration for deliberative projects. Projects should be explicit about their motivations, objectives, and leadership. Projects should be designed with the goal of building trust, and should have definite communications strategies that engage in the work of earning public trust over time.

For AI-enhanced projects, such transparency should extend into disclosure of their AI models and design decision process. We recommend projects ensure adequate disclosure of AI use and involve stakeholder participation in the development of AI use cases. This includes performing testing for unintended bias and using open-source platforms and language models to enable audits of the code and training data. This openness allows for crowd-sourced improvements and greater transparency, ensuring the ethical deployment of AI in civic technology. Good disclosures provide clear, obvious, and visible announcements of features that rely on AI integrations, such as alert boxes next to AI-generated content, as well as detailed descriptions of how AI is used and how data is processed, such as dedicated pages explaining the implementation of those same features.

We also recommend building trust by encouraging early feedback and engagement with early adopters. Soliciting and incorporating responses from users may improve functionality, but will also win buy-in from those who participate. If structured well, these initial sessions can make important inroads with community leaders and steer the design of a project towards AI integrations that will best serve the user community.

---

With respect to the practicalities of building AI integrations, both our organizations learned how crucial it is to provide LLMs with relevant context associated with queries. We both found RAG to be a very valuable approach to do this, as it provides LLMs with scalable access to detailed information that can be drawn selectively from large corpora. Moreover, we both also found it necessary to engineer the integration of additional metadata from outside sources to augment LLM queries. MAPLE extracted the text of current law to help with summarization of pending legislation, for example, and Make.org added additional background information on the CESE.

Our final recommendations center on achieving impact. Users are more likely to participate if they know their efforts are worthwhile. Governments must commit to seriously engaging with the outputs of deliberative processes. For example, deliberative assemblies in Belgium are composed of parliamentary members and randomly selected citizens, and the Belgium government has committed to engaging with resulting recommendations.[13] While participatory processes do not require implementation, they do rely on consideration. Legislatures should plan for and disclose how they will engage with growing bodies of public testimony, such as committing to whether or not they will read all testimony submitted to a committee and identifying if they will use AI technology to help synthesize large volumes of comment.

Lastly, do not let excessive caution hinder preparedness and action. It is crucial to avoid widening the gap between institutions, the public, and the private sector; automation can be a valuable ally in tackling this growing rift. Projects can start at a small scale, gaining a foothold on assistive uses of AI in a specific context while demonstrating the opportunity to apply at a wider scale. Continuous integration helps legislators and other stakeholders try the tools, build comfort and familiarity, and prepare them for systemic use. Additionally, this process helps them become effective regulators of the technology.

## **FUTURE DIRECTIONS FOR AI & LEGISLATIVE ENGAGEMENT**

As a frontier technology, our greatest strategy for deploying AI to improve democracy is by experimenting and iteratively learning from one another. Federalism, a longstanding political structure in the US and an increasingly prominent feature of European governance, aligns well with the challenges ahead. Local and state jurisdictions, the "laboratories of democracy," can run many simultaneous experiments using AI to augment our democratic processes alongside deployments on a larger scale.

In this paper, we have discussed many developed and contemplated uses of AI, but this just scratches the surface. As digital technologies like social media harm our democratic structures and our collective intelligence capacity through mechanisms including the spread of disinformation, epistemic cynicism, and techno-affective polarization (McKay & Tenove 2020) it is imperative that we pilot pro-democracy technology, identify and elevate promising tools, and build new democratic infrastructure across the four dimensions we discussed above.

However, in addition to those four dimensions, a fifth dimension is worth considering: the ability for constituents to monitor the implementation of their directives and hold lawmakers accountable. The influence of public sentiment on policy decisions often competes with the influence of other stakeholders—namely, lobbyists, corporations, and wealthy donors. In countries that restrict these entities from exerting significant political power, through either normative or legal regulation, there is greater likelihood that AI-enhanced public consensus will be meaningfully implemented by governmental bodies. However, significant barriers exist in countries like the United States, where a uniquely broad interpretation of the First Amendment prohibits lobbying and campaign finance regulations that are commonplace in many jurisdictions. The result, as found by several researchers (Giles & Page, 2022), is that "when the preferences of economic elites and the stands of organized interest groups are controlled for, the preferences of the average American appear to have only a minuscule, near-zero, statistically non-significant impact upon public policy."

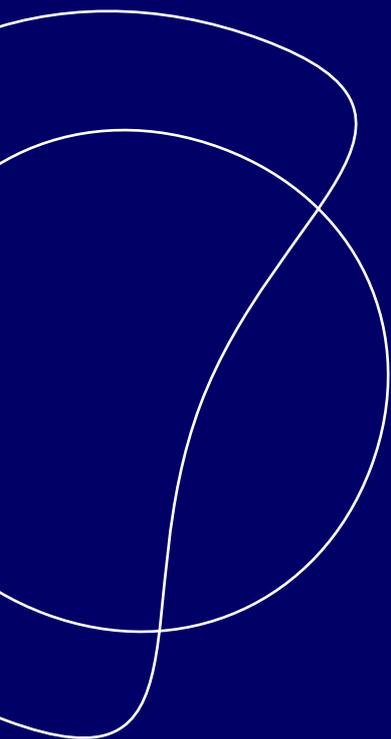

While AI deployment will not restrict the political activities of economic actors, it can radically increase our understanding of available information pertinent to monitoring elected officials, and therefore improving legislative accountability. For example, while legislator voting records and cosponsoring activities are often difficult to piece together and evaluate, AI could quickly summarize how a legislator has acted on a given issue and provide a clear picture of where a legislator stands relative to their peers. Moving from individual legislators to legislative outputs, AI can highlight discrepancies between legislative actions and the weight of public testimony or the outputs of citizen assemblies, identifying stakeholders or groups that may have exerted influence through more private channels.

Furthermore, AI can help compare a law or a project before and after citizen participation and link the changes to the input of citizens. In France, researchers have been training an AI to understand the amendments made to a law (Gesnouin et al. 2024). We can easily imagine how to extend this to tracking citizen-led changes to a text.

AI can also help the public stay informed with the development of a legislative project by making all the debates and relevant news easily accessible. Make.org illustrated this deployment during the debate on the citizen convention on end of life.[14] If an AI is fed with all the news regarding the implementation of a project, citizens will be able to ask questions to follow the progress, and then raise their concern or simply manifest their approval.

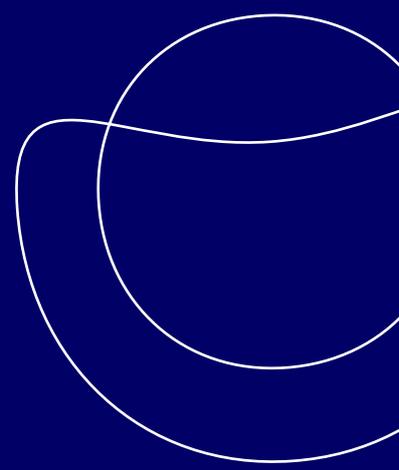

14    https://panoramic.make.org/customer/cese/event/convention-citoyenne-sur-la-fin-de-viehttps://
       panoramic.make.org/customer/cese/event/convention-citoyenne-sur-la-fin-de-vie

# CONCLUSION

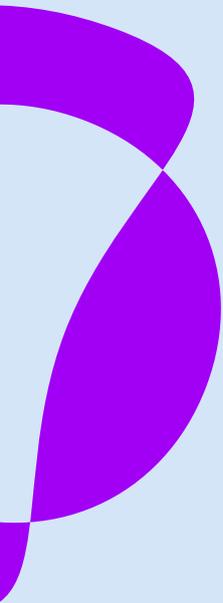

In a democracy, legislation and public policy should be created by leveraging and carefully weighing the knowledge and experience of all relevant stakeholders. This is necessary both to ensure the most beneficial and fair policy outcomes, and also to legitimize the democratic system in the eyes of the communities it serves and governs. We have described how generative AI presents profound opportunities to elicit and amalgamate the wisdom of any number of stakeholders. Through our case studies, we have sought to show how it can do so in ways that foster meaningful and dignified participation so that stakeholders recognize the output as a community-decision. This great potential will be difficult to manifest, but the authors of this paper are encouraged by the growing community dedicated to the task.

> In that context, this paper has introduced two organizations that are developing citizen participation and legislative engagement platforms integrating AI assistive features: Make.org in Europe and MAPLE in the US (Massachusetts).

We have discussed the approach each organization has taken to developing tools that serve four key dimensions of democratic process: (1) understanding complex policies, (2) helping people express their viewpoint, (3) facilitating deliberation and consensus, and (4) conveying insight to policymakers. We have compared and contrasted their projects in the context of jurisdictional differences and synthesized recommendations for civic organizations and governments.

Finally, we welcome feedback and collaboration from scholars, policymakers, and other stakeholders. Both our projects are developed in an open-source model and our code repositories can be both jumping-off points for others looking to deploy AI in legislative contexts and for those looking to propose new or critical ideas.[15]

---

15    See https://gitlab.com/makeorg and https://github.com/maple-testimony/maple

# ACKNOWLEDGMENTS

The authors thank IE University for the opportunity to develop this manuscript and Bruce Schneier and Beth Friedman for their input to and feedback on this paper. Victor and Sanders thank all members of the MAPLE team, including James Vasquez and all the volunteers with Code for Boston, the Boston University SPARK student team, Dan Jackson at NuLawLab, and all users of the platform. Sanders thanks members of the Health Resources in Action staff for feedback on test cases of AI testimony summarization. Combaz and Mas thank the CESE and the French Ministry of Economy for trusting our platform and the high-quality collaboration. They also thank the BPI, Sciences Po Paris, and the Sorbonne University for the incredible scientific partnership they provide and the Make.org team for their enthusiasm and their relentless work, without which none of this would have been possible. They also thank Dario Garcia De Viedma Ferreras, Irene Blazquez Navarro, and Carlos Luca De Tena. The authors thanks Lucas Schmuck for insightful references to related literature.

# DATA & MATERIALS

**MAPLE:** https://github.com/codeforboston/maple
**Make.org:** https://github.com/makeorg

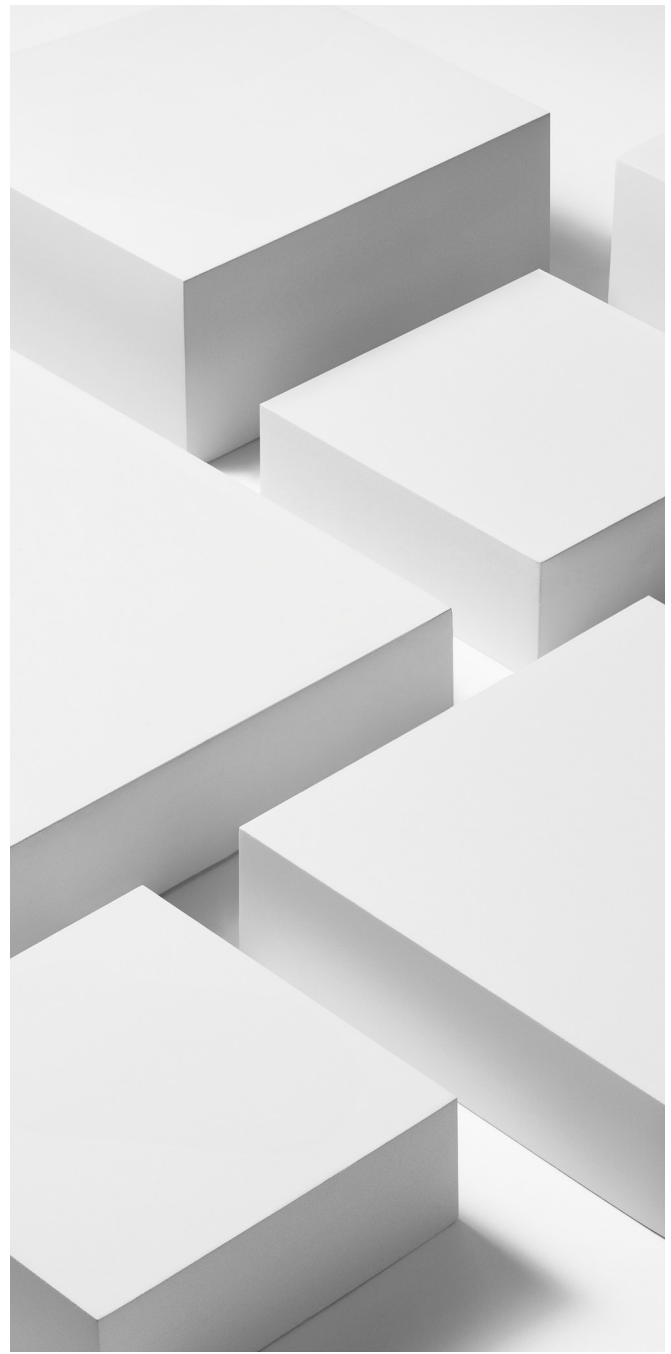

**Written by:**

Alicia Combaz, Make.org

David Mas, Make.org

Nathan Sanders, MAPLE

Matthew Victor, MAPLE

This paper is part of a series of four papers within AI4Democracy, a global research and outreach initiative led by the Center for the Governance of Change at IE University, with Microsoft as strategic supporter. AI4Democracy seeks to harness AI to defend and strengthen democracy through coalition-building, advocacy, and intellectual leadership.

**Suggested citation:**

Combaz, A., Mas, D., Sanders, N., & Victor, M. (2024) *Applications of Artificial Intelligence Tools to Enhance Legislative Engagement, AI4Democracy*, IE Center for the Governance of Change.

**FOR MORE INFORMATION ON THE AI4DEMOCRACY INITIATIVE, VISIT:**

IE.EDU/CGC/RESEARCH/AI4DEMOCRACY

cgc.ie.edu